\documentstyle[12pt,a4,epsf]{article}
\def\BR{{\rm I\!R}}
\def\BN{{\rm I\!N}}
\def\BH{{\rm I\!H}}
\def\BI{{\rm 1\!l}}
\def\BZ{{\rm Z\!\!Z}}
\def\be{\begin{equation}}
\def\ee{\end{equation}}
\def\ba{\begin{array}}
\def\ea{\end{array}}
\def\bea{\begin{eqnarray}}
\def\eea{\end{eqnarray}}
\def\bl{\begin{list}{}{}}
\def\el{\end{list}}
\def\bean{\begin{eqnarray*}}
\def\eean{\end{eqnarray*}}
\def\ts{\textstyle}
\def\ds{\displaystyle}
\def\w{{\cal W}}
\def\fermion#1{
  \begin{picture}(40,10)
    \thinlines
    \put(0,0){\vector(1,0){20}}
    \put(20,0){\line(1,0){20}}
    \put(10,2){$\scriptstyle #1$}
  \end{picture}
}
\def\boson#1{
  \begin{picture}(40,10)
    \thinlines
    \multiput(0,0)(8,0){5}{
      \begin{picture}(8,4)
        \put(0,0){\oval(4,4)[b]}
        \put(4,0){\oval(4,4)[t]}
      \end{picture}
    }
    \put(15,5){$\scriptstyle #1$}
  \end{picture}
}
\def\vertex#1#2#3{
  \begin{picture}(90,80)
    \put(40,40){\vector(-1,1){20}}
    \put(20,60){\line(-1,1){20}}
    \put(30,60){$\scriptstyle #1$}
    \put(0,0){\vector(1,1){20}}
    \put(20,20){\line(1,1){20}}
    \put(30,20){$\scriptstyle #2$}
    \put(35,40){\boson{#3}}
  \end{picture}
}
\def\graph#1#2#3#4#5#6{
  \begin{picture}(120,80)
     \put(40,40){\vector(-1,1){20}}
     \put(20,60){\line(-1,1){20}}
     \put(30,60){$\scriptstyle #1$}
     \put(0,0){\vector(1,1){20}}
     \put(20,20){\line(1,1){20}}
     \put(30,20){$\scriptstyle #2$}
     \put(35,40){\boson{#5}}
     \put(120,40){#6}
     \put(80,40){\vector(1,1){20}}
     \put(100,60){\line(1,1){20}}
     \put(110,60){$\scriptstyle #3$}
     \put(120,0){\vector(-1,1){20}}
     \put(100,20){\line(-1,1){20}}
     \put(110,20){$\scriptstyle #4$}
  \end{picture}
}
\def\pn{\par\noindent}
\def\pano{\par\noindent}
\def\Win{${\cal W}_{1+\infty}$}
\def\del{\partial}
\def\deb{\bar{\partial}}
\def\delb{\bar{\partial}}
\def\da{\dagger}

\def\ha{a^{\da}}
\def\BBCT{\,\hbox{\hbox to -2.8pt{\vrule height 6.7pt width .3pt
  \hss}\rm C}}
\def\BBCS{\,\hbox{\hbox to -2.2pt{\vrule height 4.5pt width.2pt
  \hss}$\scriptstyle\rm C$}}
\def\BBCSS{\,\hbox{\hbox to -2pt{\vrule height 3.3pt width .2pt
  \hss}$\scriptscriptstyle \rm C$}}
\def\BC{{\mathchoice{\BBCT}{\BBCT}{\BBCS}{\BBCSS}}}

\newcommand{\reseteqn}{\setcounter{equation}{0}}
\newcommand{\mysection}{\reseteqn\section}
\renewcommand{\thefootnote}{\fnsymbol{footnote}}
\begin{document}
  \pagestyle{empty}
  \begin{raggedleft}
%CSIC-IMAFF-??\\
hep-th/9412053\\
November 1994\\
  \end{raggedleft}
  $\phantom{x}$\vskip 0.618cm\par
  {\huge\bf
  \begin{center}
On a New Universal Class of\\
Phase Transitions\\
and\\
Quantum Hall Effect
  \end{center}
  }\par
  \vfill
  \begin{center}
$\phantom{X}$\\
{\Large Michael A.I.~Flohr\footnote[1]{email: iffflohr@roca.csic.es}}\\
{\em Instituto de Matem\'aticas y F\'{\i}sica Fundamental, C.S.I.C.,\\
Serrano 123, E-28006 Madrid, Spain}
  \end{center}\par
  \vfill
  \begin{abstract}
  \noindent
  We study the possible phase transitions between $(2+1)$-dimensional
  abelian
  Chern-Simons theories. We show that they may be described by non-unitary
  rational conformal field theories with $c_{{\it eff}} = 1$. As an example
  we choose the fractional quantum Hall effect and derive all its main
  features from the discrete fractal structure of the moduli space of these
  non-unitary transition conformal field theories and some large scale
  principles. Rationality of these theories is intimately related to the
  modular group yielding severe conditions on the possible phase transitions.
  This gives a natural explanation for both, the values and the widths,
  of the observed fractional Hall plateaux.
  \end{abstract}
  \vfill
  \newpage
%
%%< INTRODUCTION >%%%%%%%%%%%%%%%%%%%%%%%%%%%%%%%%%%%%%%%%%%%%%%%%%%%%%%%%
%
  \setcounter{page}{1}
  \pagestyle{plain}
  \renewcommand{\thefootnote}{\roman{footnote}}
  \mysection{Introduction}
  \pn
The quantum Hall effect (QHE) certainly is one of the most exciting
phenomena in condensed matter physics, since theory more or less fails to
describe it in terms of two- or three-dimensional continuous second order
phase transitions of the state of its electrons. The same is true for
other famous phenomena such as high temperature superconductivity.
Thus, there are condensed matter systems in reality, whose physics can be
described more or less completely in two dimensions, which show a kind
of phase transitions, but which cannot be explained by the well known
universality classes of two-dimensional statistical systems.
  \par
The aim of this paper is to introduce a new class of two-dimensional
phase transitions. As expected in two dimensions, they obey conformal
invariance at the critical point, such that the theories at the critical
point can be described by certain non-unitary rational conformal field
theories (RCFTs). This is a remarkable fact by itself, since up to now there
are not many ``useful'' applications of non-unitary theories known.
  \par
Actually, our new class of phase transition describes the interpolation
{\em between} phases of a two-dimensional system which can be described
by rational models of current algebras. In this paper we confine ourselfs
to the case of abelian current algebras, i.e.\ to the well known rational
models with $c = 1$. As it turns out, the possible transitions between two
such models are determined by severe restricitions which are deeply connected
with the modular group.
  \par
As an example, we choose the quantum Hall effect and show how the transitions
between the Hall plateaux can be described within the above mentioned new
class. Our transitions connect two different quantum Hall states by connecting
the corresponding chiral conformal theories living on their edges.
Many of the features of this class of phase transitions are very
general and automatically link together many viewpoints common in the
field of two-dimensionally confined systems such as the QHE or high-$T_c$
superconductivity. We mention only a few keywords: Anyons, Chern-Simons
theory, duality, modular group, etc.
  \par
We are able to derive all the essential macroscopic data of the QHE only
from a careful study of the nature of the new class of transitions. The
fact that the transitions are described by RCFTs imposes severe
restricitons to which states a given quantum Hall state may change, which
are intimately related to the modular group. Most
remarkebly, we find a natural explanation not only for the fact that the
Hall conductance is quantized, but also that it remains constant in some
range of variation of the magnetic field.
  \par
Allthough we mainly concentrate on the QHE, it is clear from the general
structure of our class of transitions that it should equally well addapt to
other phenomena with similar phase diagrams, such as high-$T_c$
superconductivity. More generally, we expect that these transitions appear,
whenever a system may be described by Chern-Simons theory, since the latter
is equivalent to a chiral conformal theory on the boundary of the system
in space describing a certain phase of it.
  \par
As pointed out above, we confine ourselfs in this paper to the study of
transitions belonging to abelian Chern-Simons theories. The paper is organized
as follows:
  \par
In section two we review briefly the microscopic description of the
QHE and the macroscopic large scale behaviour which is completely
governed by the Chern-Simons action. We develop a graphical description
of the possible QHE states in terms of an $1/N$ expansion of Chern-Simons-QED
Feynman graphs, form which we easily can read off the interesting observables.
  \par
In the third section we introduce some tools from conformal field theory and
give a short survey on the relation of QHE wave functions to correlators of
certain $c=1$ conformal field theories.
  \par
The fourth section proposes our scheme of phase transitions between different
QHE states, i.e.\ between different Chern-Simons theories. We argue that
certain non-unitary rational conformal theories may interpolate between
different chiral $c=1$ conformal field theories. We use the concept
of fusion well known in rational conformal theory to describe the attachment
of flux quanta to particles according to Jain's picture of the QHE. By this
procedure we move from one sector of such a non-unitary theory to another
such that chiral projection leads to different $c=1$ theories.
  \par
Section five discusses the structure of possible phase transition of the
proposed kind due to an action of the modular group on the parameters of the
discrete fractal moduli space of our non-unitary theories. Together with
the large scale principles of the first section this enables us to predict
the observable fractional values as well as the widths of their plateaux.
They correspond to certain attractor regimes in the moduli space.
  \par
In the last section we summarize our results and give some more details on
the attractor band structure. We also point out possible generalizations and
further applications of our scheme to other two-dimensional phenomena in
condensed matter physics, such as high-$T_c$ superconductivity.
%
%%< QHE -> CHERN-SIMONS >%%%%%%%%%%%%%%%%%%%%%%%%%%%%%%%%%%%%%%%%%%%%%%%%%
%
  \par
  \mysection{From Quantum Hall Effect to Chern-Simons-Theory}
  \pn
The experimental discoveries of the integer quantum Hall effect (IQHE)
\cite{Kli80} 1980 and of the fractional quantum Hall effect (FQHE)
\cite{TSG82} 1982 are one of the most interesting physical phenomena in solid
state physics in recent years \cite{PG87,Sto92}. The transversal conductance
of a two-dimensional electron gas in a high magnetic field at low temperature
exhibts quantized plateau values of the form $\sigma_{xy}={e^2 \over h}\nu$,
where the filling factor $\nu$ is an integer or fractional number. In many
respects, both the integer as well as the fractional effect share very
similar underlying physical characteristics and concepts, for instance the
two-dimensionality of the system, the quantization of the Hall conductance
with simultaneous vanishing of the longitudinal resistance, and the
interplay between disorder and the magnetic field giving rise to the
existence of extended states. In other respects, they encompass entirely
different physical principles and ideas. In particular, while the IQHE is
essentially thought of as a noninteracting electron phenomenon \cite{Lau81},
the FQHE is believed to arise from a condensation of the two-dimensional
electrons into a new incompressible state of matter as a result of
interelectron interaction \cite{Lau83}, the so called quantum droplet.
  \par
This condensation phenomenon could be extended to a whole hierarchic
structure of quasi-particles and -holes, which is based on the fundamental
states with $\nu = 1/(2p+1)$ \cite{Hld83,Hlp84}. However, the repeated
condensation of quasi-paritcles seems somehow unphysical. Another
phenomenological theory of J.K.~Jain considers composite particles built from
electrons and attached flux quanta of the magnetic field. In this model
IQHE and FQHE appear in a unified way. Recent experimental results are in
good agreement with this theory \cite{DST93,WRP93,HLR93}.
  \par
In most of the works on the FQHE, the ansatz of R.B.~Laughlin for the wave
functions to the fundamental fractions $\nu = {1 \over 3},{1 \over 5},
{1 \over 7} \dots$ plays an important r\^ole \cite{Lau83}. This ansatz cannot
yet prooven rigorously, but has an extremely high overlap with numerical
exact solutions. The ground state takes the simple form
  \begin{equation}\label{eq:laughlin}
    \psi_p = \prod_{i<j}(z_i-z_j)^{2p+1}\exp\left(-\frac{1}{2}\sum_i|z_i|^2
    \right)\,,
  \end{equation}
where $p$ should be an integer due to the Pauli principle. In J.K.~Jain's
picture, where the electrons are bound to $2p$ flux quanta, the wave functions
are obtained from the ones of the IQHE, $\phi_n$, with $\nu=n$, by
  \begin{equation}\label{eq:jain}
    \psi_{\nu} = D^{2p}\phi_n\ \ \ {\rm with}\ \ \ D = \prod_{i<j}(z_i-z_j)\,.
  \end{equation}
Mean field arguments yield the filling factor $\nu= n/(2pn\pm 1)$. The
Laughlin wave functions are obtained for $n=1$. The assumption, that an even
number of flux quanta is attached, results from the requirement that the
statistics of the composite particles remains fermionic.
  \par
The incompressibility of these quantum fluids is explained by an finite
energy gap above the ground state. This incompressibility also results in an
infinite symmetry which describes the area preserving nonsingular deformations
of the quantum droplet and commutes with the hamiltonian
\cite{CTZ93,FlVa93,Kar94}.
The quantization of this symmetry is well known in physics as the nonsingular
part of a \Win\ and arises e.g.\ in string theories or two-dimensional
gravity \cite{PRS90,Bak89,Wit92}. These deformations are directly related to
edge excitations which should live on the one-dimensional boundaries and were
studied by a number of authors \cite{Hlp82,Wen92,BW90,FrKe91,FZ91,CDT93}.
The dynamics of these edge states is mainly based on the relation
of Chern-Simons gauge theories and conformal field theory \cite{Wit83,Wit89}.
  \par
  \subsection{Microscopical Description}
  \pn
Let us consider a two-dimensional electron in a uniform transversal magnetic
field $B$. The Schr\"odinger-Equation then takes the form
  \be
    H\psi = \frac{1}{2m}\left(p-\frac{e}{c}A^2\right)^2\psi = E\psi\,,
  \ee
where the momentum $p = -i\hbar\nabla$ and the gauge potential $A$ are defined
in the plane. Let us choose the symmetric gauge ${\bf A}={B \over 2}(-y,x)$
and introduce complex variables: $z=x+iy$, $\bar z = x-iy$ and $\del =
{1\over 2}(\del_x - i \del_y)$, $\deb = {1\over 2}(\del_x + i\del_y)$.
Defining all lengths in units of the magnetic length,
  \begin{equation}
    l=\Big({2\hbar c\over eB}\Big)^{1\over 2}\,,
  \end{equation}
and the energies in units of the Landau level spacing,
  \begin{equation}
    \omega_c = {eB \over mc}\,,
  \end{equation}
the Hamiltonian can be reexpressed as:
  \begin{equation}
    H = 2\hbar\omega_c l^{2}\big(-\del\deb + {1\over 2l^2}(\bar z\deb - z\del)
    + {1\over 4l^4}z\bar z \big)\,.
  \end{equation}
Letting $\hbar=m=l=1$ the hamiltonian and the angular momentum $J$ can be
written in terms of a pair of independent harmonic oscillators:
  \begin{eqnarray}
    H &=& a^{\da}a + aa^{\da}\,,\\
    J &=& b^{\da}b - a^{\da}a\,,
  \end{eqnarray}
where these operators are
  \begin{eqnarray}
    a={z\over 2} + \deb\,,&\qquad\qquad &a^{\da} = {\bar z\over 2} - \del\,,\\
    b={\bar z\over 2} + \del\,,&\qquad\qquad &b^{\da} = {z\over 2} - \deb\,,
  \end{eqnarray}
and satisfy the canonical commutation relations
  \begin{equation}
    [a,a^{\da}] = 1\,, \qquad\qquad [b,b^{\da}] = 1\,,
  \end{equation}
with all other commutators vanishing. The vacuum is defined by the condition
$a\psi_{0,0}=b\psi_{0,0}=0$ which yields
  \be
    \psi_{0,0} = \frac{1}{\sqrt{\pi}}\exp(\frac{1}{2}|z|^2)\,.
  \ee
The solutions of the Schr\"odimger-Equation are divided into infinitely
degenerated Landau levels (due to rotational invariance around the axis of
the magnetic field) with energies $2n+1$ and eigen functions
  \be
    \psi_{n,l} = \frac{(b^{\dagger})^l(a^{\dagger})^n}{\sqrt{l!n!}}\psi_{0,0}
    \,.
  \ee
Restricting the model to finite size of area $A$ reduces the degeneracy, since
higher angular momentums yield wave functions with larger support. The
degeneracy of each Landau level is then given as $N_A = \Phi_{{\em mag}}/
\Phi_0$, where $\Phi_{{\em mag}} = BA$ is the magnetic Flux through the plane
and $\Phi_0=(h/e)$ is the elementary Flux quantum.
  \par
Suppose now, there are $N$ such electrons. If there is no interaction between
them, then only the magnetic field $B$ controls the number of states and thus
the density of electrons per state, acting as an external pressure. The
many-particle problem splits in $N$ copies of the single-particle problem with
identical operators $a_i = a, b_i = b$, where the label $i$ refers to the
$i$-th electron. Since the electron density per state is the correct quantum
measure of the electron density, i.e.\ the filling fraction $\nu$, the latter
is given as $\nu=N/N_A$ and is forced to be an integer due to certain gauge
conditions. In fact, $\nu$ can be viewed as the Chern-character of an U(1)
line bundle over the parameter torus of the magnetic fluxes, and thus is an
integer valued topological invariant \cite{NTW85,Koh85,AS85,Nov81}. It also
may be related to an element in the cyclic cohomology of a $C^*$ algebra
\cite{Bel86}.
  \par
Let us now introduce a Chern-Simons type interaction of flux quanta that in
Jain's picture are thought of being attached to the electrons. To that issue
we redefine the operators $b_i, a^{\dagger}_i$,
  \be
    \begin{array}{rcl}
      b_i&=&\partial_i+\frac{\bar z_i}{2} - 2p\sum_{i\neq j}\frac{1}{z_i-z_j}
      \,,\\
      a_i^{\dagger}&=&-\partial_i+\frac{\bar z_i}{2} + 2p\sum_{i\neq j}
      \frac{1}{z_i-z_j}\,.
    \end{array}
  \ee
Of course, we now get additional terms of the form $2p\pi\delta(z_i-z_j)$
in the commutation relations,
  \begin{eqnarray}
    [a_i,a_i^{\da}] &=& 1 - 2p\pi \sum_{i\neq j} \delta(z_i - z_j)\,,\\
    \ [a_i,a_j^{\da}] &=& 2p\pi \delta(z_i - z_j)\ \ {\rm for}\ \ i\neq j\,.
  \end{eqnarray}
But these terms may be neglected as long as we
require fermionic statistics, i.e.\ vanishing of the wave functions, if
two paricle coordinates approach each other. The resulting theory describes
the fractional quantum Hall effect with filling $\nu=1/(2p+1)$, where the
macroscopic observables are obtained by addition as in the case without
interaction, $H = \sum_{i=1}^N(a_i^{\dagger}a_i+a_ia_i^{\dagger})$ and the
Laughlin wave function (\ref{eq:laughlin}) is eigen function of $H$ to the
lowest Landau level.
  \par
As has been widely pointed out in the literature (see e.g.\
\cite{CTZ93,Kar94,FlVa93}),
the incompressibility of the Laughlin quantum droplet is equivalent
with the invariance of the theory under non singular area preserving
diffeomorphisms. With our definitions from above and $n,m\geq -1$ we can
define operators
  \be
    {\cal L}_{m,n} = \sum_{i=1}^N(b_i^{\dagger})^{m+1}(b_i)^{n+1}\,,
  \ee
which all commute with the Hamiltonian\footnote{Note that there even do not
appear any terms with $\delta$-functions.}. They generate the algebra
$\w_{1+\infty}^+ = \w^+(1,2,3,4,\ldots)$ of non-singular area preserving
diffeomorphisms with commutation relations
  \be
    [{\cal L}_{n,m},{\cal L}_{k,l}]=\sum_{s=0}^{{\rm min}(m,k)}
    \frac{(m+1)!(k+1)!}{(m-s)!(j-s)!(s+1)!}{\cal L}_{n+k-s,m+l-s}
    -(m\leftrightarrow l,n\leftrightarrow k)\,.
  \ee
Moreover, the Laughlin wave functions $\psi_p$ from (\ref{eq:laughlin})
are lowest-weight states, i.e.\ with $W_n^{(s)}\sim{\cal L}_{n+s-2,s-2}$
for $s\geq 1$, $n\geq -s+1$ as the fourier modes of the generators of the
algebra with spin $s$ we have $W_n^{(s)}\psi_p = 0$ for $-s<n\leq -1$.
  \par
Let us make a few remarks: The redefined operators $b_i$ and $a_i^{\dagger}$
for $p>0$ do not longer have $b_i^{\dagger}$ and $a_i$ as their Hermitian
adjoints. This can be corrected by introducing an inner product
  \begin{equation}\label{eq:sp}
    \left\langle\Psi_1 \mid \Psi_2\right\rangle = \int \Psi_1^{\da} \mu
    \Psi_2^{}\,,
  \end{equation}
where the non trivial singular measure $\mu$ is given as:
  \begin{equation}\label{eq:mu}
    \mu(z_1,\bar z_1,\dots,z_N,\bar z_N) = \prod_{i<j}
    \mid z_i - z_j \mid^{-4p}\,.
  \end{equation}
We may observe that the Laughlin wave functions are also eigen functions to
the free Hamiltonian of the original $a_i, a_i^{\dagger}$ operators with the
same eigen value. Thus, the Chern-Simons interaction does not destroy the
Landau level structure.
  \par
The configuration space for distinguishable particles is given by
  \begin{equation}
    {\rm C}_N = \{(z_1,\dots,z_N) \in \BC^N; \quad z_i \neq z_j
    \quad {\rm for} \quad i \neq j \} \,.
  \end{equation}
The $(a_i,\ha_i)$ can be considered as covariant derivatives on a $U(1)
\otimes \dots \otimes U(1)$ bundle over C$_N$ as in the paper
of E.~Verlinde on the non-abelian Aharanov-Bohm effect \cite{Ver91}.
Thus, the curvature is given by (\ref{eq:mu}) which describes a constant
magnetic field plus $2p$ flux quanta added to each electron. This is exactly
the FQHE interpretation of J.K.~Jain by the composite fermion theory mentioned
above. These flux quanta can be described in an abelian Chern-Simons theory by
localized Wilson loops. Considering the $N$-point function of these flux
quanta localized at the positions $z_i$ of the electrons one sees that it is
proportional to the measure $\mu$ (\ref{eq:mu}) using that those Wilson
loop operators can be expressed by vertex operators \cite{BBGS92}.
This explains the former observation on the relation between vertex operator
correlators and the Laughlin wave function \cite{Fub91,Sto91,CMM91,MR90}.
  \par
This picture is in good agreement with an argument of A.~Lopez and E.~Fradkin
\cite{FL90} that adding an even number of flux quanta to
each electron leaves all expectation values invariant. One can see this, if
one calculates expectation values via path integrals, because then only
closed paths contribute to the partition function which is ${tr}\,\exp(
-\beta H)$. Closed paths are given by exchanging electrons or moving them
around each other. The phase associated to a path loop has two contributions,
the statistics of the particles and the Aharanov-Bohm phase due to the
enclosed flux. The former is fermionic, since adding an even number of flux
quanta to the electrons does not change it, the latter is trivial, since
each whole flux quantum produces a phase of unity. Calculating the expectation
values of the Laughlin wave function with the inner product (\ref{eq:sp}),
  \begin{equation}
    \int \psi_p^{\da}\,\mu\, \psi_p^{\phantom{\da}} dz^N \,,
  \end{equation}
it is easy to see that this expression is independent of $p$, thus, adding
flux quanta indeed does not change the expectation value.
  \par
Thus, in our formulation of the FQHE we consider a Hamiltonian without
explicit interelectron interaction as in the IQHE, but describing the
interaction with the help of a nontrivial measure coming from the $N$-point
correlation function of the flux quanta in an abelian Chern-Simons theory.
  \par
Let us emphasize again that this picture of the FQHE is not only a complicated
view of the IQHE. Since adding of two flux quanta does change the effective
magnetic field and thus, the size of the wave functions, not all expectation
values remain unchanged. Moreover, E.~Verlinde \cite{Ver91} has shown that
a free Hamiltonian (without magnetic field) acquires a non trivial
$S$-matrix when a substitution of the kind
  \be
    \del_i \rightarrow \del_i + \alpha\sum_{i\neq j}\frac{1}{z_i-z_j}\,,
    \quad\quad\delb_i \rightarrow \delb_i
  \ee
is applied in the covariant derivatives. On the other hand, a Hamiltonian
with Chern-Simons interaction can be rewritten as a free Hamiltonian of
non interacting particles with anyonic statistics via a singular gauge
transformation eliminating the Chern-Simons gauge field. The particles still
are subject to a non trivial $S$-matrix.
Both these aspect will be very important in the following.
  \par
We conclude our short introduction to the microscopic aspects of the
QHE by mentioning that one can easily generalize the Chern-Simons interaction
terms to the case of different independent quantum fluids (i.e.\ sets of
eventually interacting Landau levels or different layers), see e.g.\
\cite{Wen92,Sto91,WZ92,Jai89b,FZ91,FlVa93}.
Such systems lead to Laughlin type wave functions of
the form
  \be\label{eq:LT}
    \psi_K(\{z_i^I\}) \sim \prod_{{I\atop i<j}}(z_i^I - z_j^I)^{K_{II}}
    \prod_{{I<J\atop i\leq j}}(z_i^I - z_j^J)^{K_{IJ}}\exp(-\frac{1}{4l^2}
    \sum_{I,i}\mid z_i^I\mid^2)\,,
  \ee
where now the electrons are distributed to $n$ different subbands\footnote{
Subbands can be realized as different layers, different Landau levels or
additional quantum numbers for the first Landau level.}
which are labeled by $I,J = 1,\ldots,n$, each subband $I$ containing $N_I$
electrons. The Pauli principle and the requirement of single valuedness of the
wave function restrict the $n\times n$ matrix $K_{IJ}$ to be symmetric,
positive, integer valued with odd integers on the main diagonal.
The filling fraction of such states is given by
  \be\label{eq:Knu}
    \nu = \sum_{IJ}(K^{-1})_{IJ}\,.
  \ee
The filling fractions of J.K.~Jain's model are obtained from the simple
$n\times n$ matrices $K = \BI_n + 2pC_n$, where $C_n$ is the $n\times n$
matrix with all entries equal to 1. The resulting Lagrangian describes a
system of $n$ independent currents which, however, are coupled together
by just the global Chern-Simons coupling via $2p$ flux quanta.
  \par
The fractional fillings can also be understood in terms of topological
invariants, namely the first Chern-number of a vector bundle divided by its
rank, where the latter is equal to the degeneracy of the ground state
\cite{NTW85}. For a recent treatment which relates the observed fractions
to the class of stable vector bundles see \cite{Var94}.
  \par
  \subsection{Macroscopic Description}
  \pn
Let as again consider a system of non interacting electrons in a strong
transverse magnetic field $B$, confined to a two-dimensional domain (thus,
we live in $(2+1)$-dimensional space time). The current $J^{\mu}$ can be
written as the curl of a vector potential $\alpha$, i.e.
  \be
    J^{\mu} = \frac{1}{2\pi}\epsilon^{\mu\nu\lambda}\del_{\nu}
    \alpha_{\lambda}\,.
  \ee
Since $J^{\mu}$ is invariant under gauge transformations of $\alpha$,
$\alpha_{\mu}$ is a gauge field.
By gauge invariance, the simplest possible local term in the Lagrangian
density is just the Chern-Simons term, i.e.
  \be\label{eq:csl}
    {\cal L} = \frac{\eta}{4\pi}\epsilon^{\mu\nu\lambda}
    \alpha_{\mu}\del_{\nu}\alpha_{\lambda} + \ldots \,,
  \ee
where $\eta$ is a dimensionless coefficient. As has been discussed by
J.~Fr\"ohlich and A.~Zee \cite{FZ91}, there could be other terms including
the Maxwell term $\frac{1}{g^2}f_{\mu\nu}^2$ and other short-range dynamical
interactions. But these additional terms will be invisible in the scaling
limit, if $\eta\neq 0$. Actually, as argued by them, every $(2+1)$-dimensional
gauge theory at zero temperature with a strictly positive energy gap will be
completly governed at very long distances by the topological
Dirac-Aharanov-Bohm type interactions between charged sources carrying
magnetic vorticity. These topological interactions are described by a
Chern-Simons Lagrangian as given by (\ref{eq:csl}).
  \par
In the simplest case of one filled Landau level, it turns out that
$\eta = \nu = 1$.
Of course, we could take different currents $J_I$ and we could introduce
electron-electron interactions. The universality of the long distance
behaviour, however, forces that the only effect of electron-electron
interactions is to modify the coefficient $\eta$, as long as the Landau
level structure and the positive energy gap are preserved. Thus the
effective long distance Lagrangian is just given as
  \be\label{eq:csll}
    {\cal L} = \frac{1}{4\pi}\sum_{I,J}K_{IJ}\epsilon^{\mu\nu\lambda}
    \alpha_{I,\mu}\del_{\nu}\alpha_{J,\lambda} + \ldots \,,
  \ee
with the same matrix $K_{IJ}$ as introduced at the end of the last section.
This way of describing the QHE with abelian Chern-Simons theory has been
extensively studied \cite{FrKe91,FZ91,FL90}.
  \par
There is a second principle, which further restricts the influence of the
microscopic behaviour to the macroscopic observables: As has been pointed
out in the last section, we can always write the $N$-particle Hamiltonian
in an form which exhibits $O(N)$-symmetry, eventually moving non trivial
effects to some anyonic behaviour of the particles. But $O(N)$-invariant
Hamiltonians admit an $1/N$-expansion of the correlation functions. Since
$N\gg 1$, only the leading terms will survive. As we will see later, this
can explain the independence of observables, such as the transversal
conductivity within a plateau, from the variation of the external parameters,
such as the external magnetic field. In fact, a small variation leads to
microscopical influences which, however, do only contribute to higher
terms of the $1/N$-expansion.
  \par
To see this in more detail, let us view {\em abelian} Chern-Simons theory as
an ordinary quantum electrodynamics (QED) with massive bosons. Thus, we can
introduce Feynman graphs: A current of $n$ non interacting Landau levels (or
sublevels) is depicted by a fermionic line \fermion{n}\ whose direction
corresponds to the moving direction of the electrons. The Chern-Simons
interaction of $2p$ attached flux quanta is depicted by a bosonic line
(the ``photon'') \boson{1/2p}\ . This syntactically makes sense, since the
charge carriers are fermions and the attachment of an even number of flux
quanta does not change the statistics of the charge carriers, hence is of
bosonic character.
  \par
Let us assume that we consider the Feynman graphs of first order in the
$1/N$-expansion, for example a graph with two free vertices corresponding
to a two-point correlation function. But what does this correlation function
measure? In fact, the macroscopic current-current correlation of the $N$
electrons. But this is directly related to the conductivity, see e.g.\
\cite{FrKe91}. Moreover, if the contributing graphs of the $1/N$-expansion
deal with macroscopic values as the currents, they must also satisfy the
ordinary laws of classical, macroscopic electrodynamics, such as the
Kirchhoff rules. This can be translated into a particularly nice graphical
description. The $1/N$-two-point-graphs may be viewed as networks of certain
conductivities driven by certain currents generated by magnetic flux through
closed loops of the graph. We may measure the conductivity between the
two free vertices. A fermionic loop of a current of $n$ Landau levels has
conductivity $n$, the boson has conductivity $1/2p$. Note, that conductivity
and conductance are the same in our situation, hence the networks are
topological in the sense that the overall conductivity does not depend on
the positions of the vertices. This also mirrors the fact that Chern-Simons
theory is topological.
  \par
The case $p=0$ corresponds to a shortcut, since a conductor in series, which
has infinite conductivity, does not change the overall conductivity.
The bosonic propagator reduces to a point interaction with bosons of
infinite mass. This is exactly, what we should expect, since $1/N$-expansion
is meaningless in the IQHE because there is no interaction between the
particles. The macroscopic observables are just given by summing up the
$N$ identical contributions of the single-particle theory. But the graph of
first order of the single-particle theory is nothing else than a
point-interaction. In general, the elementary first order graph is
  \begin{equation}\label{eq:graph}
    \graph{}{0}{}{n}{1/2p}{.}
  \end{equation}
Since we normally measure two-point correlations, we have to close two of
the external lines to a loop. Keeping one line fixed as the incoming line,
there are up to symmetry two possibilities to connect two lines. The first is
  \begin{equation}\label{eq:sxy}
    \begin{picture}(280,80)
      \put(0,0){\graph{\sigma}{0}{n}{n}{1/2p}{$\ \ \mapsto$}}
      \put(150,0){\vertex{\sigma}{0}{1/2p}}
      \put(255,40){\circle{40}}
      \put(255,60){\vector(1,0){1}}
      \put(252,39){$\scriptstyle n$}
      \put(280,40){,}
    \end{picture}
  \end{equation}
which gives the transversal conductivity $\sigma_{xy} = n/(2pn+1)$ to first
order. If we change the direction of the magnetic field which drives the
current in the loop denoted $n$, we get $\sigma_{xy} = n/(2pn-1)$ by formally
replacing $n \mapsto -n$.
  \par
The longitudinal conductivity of a Hall sample near zero temperature is
essentially zero. There may by corrections coming from the Chern-Simons
Lagrangian, but the only possible contributions are self-energy terms.
We get the self-energy to first order by the second possibility of connecting
two lines,
  \be\label{eq:sxx}
    \begin{picture}(120,80)
      \put(0,0){\vector(1,1){20}}
      \put(20,20){\line(1,1){20}}
      \put(30,20){$\scriptstyle 0$}
      \put(35,40){\boson{1/2p}}
      \put(120,40){.}
      \put(80,40,0){\vector(1,-1){20}}
      \put(100,20){\line(1,-1){20}}
      \put(110,20){$\scriptstyle \sigma$}
      \put(60,40){\oval(40,40)[t]}
      \put(60,60){\vector(1,0){1}}
      \put(59,70){$\scriptstyle 0$}
    \end{picture}
  \ee
This graph gives a vanishing contribution, as is observed in experiments,
since no non-zero conductivity, which is driven by a current from a loop
enclosing magnetic flux enters the graph.
  \par
It is easy to see that there are 15 different possibilities to get second
order contributions to two-point correlators. But since we have to keep in
mind the directions of the currents, we see that many terms cancel, because
they contribute with opposite signs. There again only two graphs survive,
which are not cuttable, i.e.\ not separable into two disconnected subgraphs
by cutting one fermionic line. One of them gives the second order transversal
conductivity, the other is a self-energy contribution corrected by a
first-order Hall current. The two graphs are
  \begin{equation}\label{eq:sxy2}
    \begin{picture}(211,80)
      \put(0,0){\vertex{\sigma}{0}{1/2q}}
      \put(105,40){\circle{40}}
      \put(105,60){\vector(1,0){1}}
      \put(102,39){$\scriptstyle m$}
      \put(105,20){\vector(-1,0){1}}
      \put(95,10){$\scriptstyle m+\sigma_1$}
      \put(120,40){\boson{1/2p}}
      \put(186,40){\circle{40}}
      \put(186,60){\vector(1,0){1}}
      \put(183,39){$\scriptstyle n$}
      \put(211,40){,}
    \end{picture}
  \end{equation}
and
  \be\label{eq:sxx2}
    \begin{picture}(127,80)
      \put(40,40){\vector(-1,1){20}}
      \put(20,60){\line(-1,1){20}}
      \put(30,60){$\scriptstyle\sigma$}
      \put(0,0){\vector(1,1){20}}
      \put(20,20){\line(1,1){20}}
      \put(30,20){$\scriptstyle 0$}
      \put(35,40){\boson{1/2p}}
      \multiput(7,62)(0,-8){7}{
        \begin{picture}(4,8)
          \put(0,0){\oval(4,4)[l]}
          \put(0,4){\oval(4,4)[r]}
        \end{picture}
      }
      \put(-13,40){$\scriptstyle 1/2q$}
      \put(103,40){\circle{40}}
      \put(103,60){\vector(1,0){1}}
      \put(101,39){$\scriptstyle n$}
      \put(127,40){.}
    \end{picture}
  \ee
Remarkebly, the longitudinal conductivity does not vanish to second order.
One may think of this correction that the longitudinal current to second order
somehow ``sees'' the Hall current of first order.
In fact, there are mesurements of Hall conductivities,
which cannot be obtained in first order and indeed have non-vanishing
longitudinal resistance. Presumably, they are driven by a lower number of
electrons such that second order effects become visible. A possible
explanation of these Hall conductivities is given by J.~Fr\"ohlich et.\ al.\
\cite{FST94} by a classification of quantum Hall fluids.
  \par
A futher remark is necessary here. Whether a Lagrangian of form (\ref{eq:csll})
has
higher order terms in the $1/N$-expansion, depends on the form of the matrix
$K$. Viewing $K$ as a network incidence matrix, it is easy to derive the
corresponding graph which will correctly yield $\nu$ as given in
(\ref{eq:Knu}). This also determines the highest order (of loops) which
will contribute to the theory, since the Chern-Simons theory for one global
current essentially is a one-loop theory.
  \par
Every symmetric, positive, integer valued matrix $K$ with odd integer entries
on the main diagonal can be built up by a repeated procedure of
globally attaching even numbers of flux quanta and extending the matrix by
adding new currents (eventually using negative numbers). Therefore, the
graph determining the filling fraction is just a further extended version of
(\ref{eq:sxy2}). But note, that the Hall conductivity will be given just by
the first order contribution in most of the cases.
  \par
Our discussion may be generalized to $n$-point-graphs, but one has to ask in
which way one could implement a physical measurement of these expectation
values.
  \par
We have now the following macroscopic picture: The universal behaviour of the
large scale physics and its topological nature explain the fact that the
Hall conductivity is quanitzed. The fact, that the macroscopic observables
are given by the $N\rightarrow\infty$ limit of the microscopic description,
may explain that the quantization is stable against small variations of the
external magnetic field. This will become more clear in the following.
%
%%< QHE -> RCFT >%%%%%%%%%%%%%%%%%%%%%%%%%%%%%%%%%%%%%%%%%%%%%%%%%%%%%%%%%
%
  \par
  \mysection{From Quantum Hall Effect to Conformal Field Theory}
  \pn
It is by now a well known fact that $(2+1)$-dimensional Chern-Simons theory
is equivalent to $(1+1)$-dimensional chiral rational conformal field theory
living on the boundary of the space domain \cite{Wit83,Wit89}. Much work has
been done to evaluate this connection in the case of the QHE
\cite{MR90,Fub91,CMM91}.
In particular, the Laughlin wave functions could be expressed as
$N$-point correlation functions of certain vertex operators of rational
$c=1$ Gaussian models. This also explained in a nice way the occurence of
non abelian statistics and anyons.
  \par
Usually, the conformal field theory (CFT) lives on the cylinder made out
of the
the edge of the quantum droplet times a time axis. But if one wants to relate
Laughlin wave functions to chiral conformal blocks of the CFT, one has to
consider an appropriate analytical continuation back into the plane (a
Wick-rotation), which we will implicitly assume in the following.
The following statements can be found in the literature:\\
$\bullet$ The CFTs for the so called abelian QHE states, which correspond to
their abelian Chern-Simons theories, have $c = 1$;\\
$\bullet$ The attached flux quanta can be described by vertex operators, which
correspond to the localized Wilson loops of the Chern-Simons theory;\\
$\bullet$ The wave functions are then given by the correlation functions of
the vertex operators;\\
$\bullet$ There is a principle of chirality at least for the FQHE at
fillings $\nu = 1/(2p+1)$, i.e.\ the wave functions are essentially given by
the chiral conformal blocks;
  \par
  \subsection{Preliminaries in CFT}
  \pn
To be more specific, let us consider vertex operators of a free field
construction of a CFT with central charge $1 - 24\alpha_0^2$. The fourier
modes of the current $j = \del\phi$ of a scalar free field $\phi(z)$ obey
the U(1)-Kac-Moody algebra
  \be
     [j_m,j_n] = n\delta_{m+n,0}\,,
  \ee
which is known to describe the chiral edge waves, i.e.\ the energy gapless
excitations of the QHE states. The irreducible lowest-weight representations
are realized as Fock spaces ${\cal F}_{\alpha,\alpha_0}$ over the
lowest-weight states $\mid\alpha,\alpha_0\rangle$ with
  \be
    j_n\mid\alpha,\alpha_0\rangle = 0\ \ \forall n<0\,,\quad\quad
    j_0\mid\alpha,\alpha_0\rangle = \sqrt{2}\alpha\mid\alpha,\alpha_0\rangle
    \,.
  \ee
These Fock spaces ${\cal F}_{\alpha,\alpha_0}$ carry the structure of
Virasoro modules, if the Virasoro field is defined by
  \be
    L(z) = {\cal N}(j,j)(z) + \sqrt{2}\alpha_0\del_z j(z)\,,
  \ee
where ${\cal N}$ means normal ordering. The Virasoro algebra has then the
central charge $c = 1 - 24\alpha_0^2$. The lowest-weight states of the
$\widehat{\rm U(1)}$ algebra become Virasoro lowest-weight states,
  \be
    L_n\mid\alpha,\alpha_0\rangle = 0\ \ \forall n<0\,,\quad\quad
    L_0\mid\alpha,\alpha_0\rangle = h(\alpha)\mid\alpha,\alpha_0\rangle\,,
  \ee
where the conformal weight is given by $h(\alpha) = \alpha^2
- 2\alpha\alpha_0$. Finally, the vertex operators map Fock spaces into
each other, $\psi_{\alpha}:{\cal F}_{\beta,\alpha_0}\mapsto
{\cal F}_{\alpha+\beta,\alpha_0}$. Their explicit form is
  \be\label{eq:Vop}
    \psi_{\alpha}=\exp\left(-\sum_{n>0}\sqrt{2}\alpha j_n\frac{z^n}{n}\right)
                  \exp\left(-\sum_{n<0}\sqrt{2}\alpha j_n\frac{z^n}{n}\right)
                  c(\alpha)z^{-\sqrt{2}\alpha\alpha_0}\,,
  \ee
where $c(\alpha)$ commutes with all $j_n, n\neq 0$ and maps lowest-weight
states into lowest-weight states. Products of vertex operators are only
defined for radial ordered coordinates, i.e.\ $\psi_{\alpha}(z_1)
\psi_{\beta}(z_2)$ is only defined for $\mid z_1\mid > \mid z_2\mid$. The
other half is obtained by analytic continuation, where the non trivial
statistics of vertex operators shows up, $\psi_{\alpha}(z_1)\psi_{\beta}(z_2)
= \exp(2\pi i\alpha\beta)\psi_{\beta}(z_2)\psi_{\alpha}(z_1)$.
  \par
{}From this construction we can in particular obtain the $c=1$ Gaussian
models, i.e.\ the U(1)-theory of mappings of the unit circle onto a circle
of radius $R$. We choose our free field $\phi(z)$ to be compactified on a
circle with radius $R$. The partition function is then
  \be\label{eq:zgauss}
    Z(R) = \mid\eta(\tau)\mid^{-2}\sum_{(p,\bar p)\in\Gamma_R}
    q^{\frac{1}{2}p^2}\bar q^{\frac{1}{2}\bar p^2}\,,
  \ee
where $\tau$ is the modular parameter of the torus, $q = \exp(2\pi i\tau)$,
and $\eta(\tau) = q^{-1/24}\prod_{n>0}(1-q^n)$ is the Dedekind function.
The summation of the ``momenta'' is over the lattice
  \be\label{eq:lattice}
    \Gamma_R = \left\{\left.(p,\bar p) = \left(\frac{n}{R} +
    {\ts\frac{1}{2}}mR, \frac{n}{R} - {\ts\frac{1}{2}}mR\right)\right|
    n,m\in\BZ\right\}\,,
  \ee
which is self-dual, if we adopt a Lorentzian metric. The self-duality assures
that $Z(R)$ is modular invariant. The normalized vertex operators are given by
  \be\label{eq:vopgauss}
    \ba{rcl}
      V_{nm}^+(z,\bar z) =
      \sqrt{2}\cos\left[p\phi(z)+\bar p\bar{\phi}(\bar z)\right]\,,
      V_{nm}^-(z,\bar z) =
      \sqrt{2}\sin\left[p\phi(z)+\bar p\bar{\phi}(\bar z)\right]\,,
    \ea
  \ee
where the relation of $(p,\bar p)$ and $(n,m)$ is defined by the lattice
$\Gamma_R$. The combinations $V_{nm}^++iV_{nm}^-$ create states with
momentum $\pm\frac{1}{2}(p+\bar p)$ and winding number $\pm(p-\bar p)$.
  \par
The models described above yield RCFTs, whenever $2R^2 = p/q,\ p,q\in\BN$.
In these cases, $Z(R)$ can be written as a finite bilinear form of the
characters of the underlying RCFT. The latter are of the form $\chi_{\lambda}
= \Theta_{\lambda,k}/\eta$, where the elliptic functions are
$\Theta_{\lambda,k}(\tau) = \sum_{n\in\BZ}q^{(2kn+\lambda)^2/4k}$. The
partition function for $2R^2 = p/q$ can then be written as
  \be\label{eq:zdecomp}
    Z(R) = \frac{1}{\eta(\tau)\eta(\bar{\tau})}\sum_{\lambda\,{\rm mod}\,2pq}
    \Theta_{\lambda,pq}(\tau)\Theta_{\lambda',pq}(\bar{\tau})\,,
  \ee
where with $\lambda = nq+mp$ mod $2pq$ we have $\lambda' = nq-mp$ mod $2pq$.
In the following we will use a slightly different notation
$Z[2R^2] \equiv Z(R)$ for convenience. The duality of the Gaussian partition
function takes a particular simple form in this notation, $Z[x] = Z[1/x]$
for every $x\in\BR_+$.
  \par
  \subsection{Wave Functions as CFT Correlators}
  \pn
Let us consider a generic correlation function of the chiral vertex operators
(\ref{eq:Vop}) on the plane (on a genus zero Riemannian manifold). We have
the well known result
  \begin{equation}\label{eq:vev}
    \langle\Omega_{-\alpha_0}^*,\psi_{\alpha_1}(z_1)\ldots\psi_{\alpha_N}(z_N)
    \Omega_0\rangle = \prod_{i<j}(z_i-z_j)^{2\alpha_i\alpha_j}\,,
  \end{equation}
if $|z_1|>\ldots>|z_N|$ and $\sum_{i=1}^N\alpha_{i} = \alpha_0$, where
$\alpha_0$ denotes the background charge and $\Omega_{\alpha}$ the ground
state to the superselection sector of charge $\alpha$.
  \par
Since this is purely holomorphic, we cannot reproduce the non-holomorphic
parts $\exp(-1/2\sum_i\mid z_i\mid^2)$ of the Laughlin wave functions
(\ref{eq:laughlin}). Either we include this term explicitly in the
integral measure $\mu$, or we insert a term $\exp(-i\alpha\int d^2z'$
$\bar{\rho}\phi(z'))$ into the correlator (\ref{eq:vev}), where $\phi$
is again the free field and $\bar{\rho}$ is an avaraged density
$(\pi\alpha^2)^{-1}$. If one integrates this term over a disk of area
$2\pi\alpha^2N$, then the real part correctly yields the desired
exponential term for $N$ electrons, while the imaginary part contributes
a singular phase. The latter can be eliminated by an also singular
gauge transformation corresponding to the uniform external magnetic field
\cite{MR90}. In the following we will often neglect the exponential term
and absorb the external magnetic field in $\Omega_{-\alpha_0}^*(N)$,
since the integral also modifies the background charge.
  \par
Let us consider the chiral $c=1$ RCFT with compactification radius
$R^2 = 2p+1$. It can be shown \cite{MR90,Fub91,CMM91} that the vertex
operators $\psi_{\alpha}$ with $\alpha=\sqrt{2p+1}$ exactly reproduce the
the Laughlin wave functions,
  \begin{equation}\label{eq:myvev0}
    \langle\Omega_{-\alpha_0}^*(N),\prod_{i=1}^N\psi_{\sqrt{2p+1}}(z_i)
    \Omega_0\rangle = \prod_{i<j}(z_i-z_j)^{2p+1}\exp\left(-{\ts\frac{1}{2}}
    \sum_i\mid z_i\mid^2\right)\,,
  \end{equation}
while the other fundamental vertex operators with charges $\alpha =
\lambda/\sqrt{2p+1}$, $\lambda = 1,\ldots,2p+1$, produce excitations with
anyonic statistics $\theta = \pi\alpha^2$. From (\ref{eq:lattice}) we learn
that the chiral vertex operators have electric charge $\alpha/R = \lambda/R^2$
and magnetic vorticity $\alpha R = \lambda$. Thus, our vertex operators
$\psi_{\sqrt{2p+1}}(z)$ have charge 1 and vorticity $2p+1$ which is, what we
would expect from composite fermions with $2p$ attached flux quanta.
  \par
There is a broad discussion in the literature on the properties of the CFT
picture of the QHE. A particular nice point is, that one can easily obtain
the wave functions for arbitrary (periodic) boundary conditions and arbitrary
genus Riemannian manifolds. In fact, the real physical system has much of a
torus, since one has to close the circuits in order to measure currents or
voltages. If one thinks of the longitudinal current generated by a magnetic
field
and measures the Hall voltage by an induced magnetic flux, one gets the so
called magnetic torus via gauge invariance modulo flux quanta. If the torus
has the modular parameter $\tau$ with (complex) lengths $L_x,L_y$, then
the generic $N$-point correlator is
  \begin{eqnarray}\label{eq:vevtorus}\lefteqn{%
    \langle\Omega_{-\alpha_0}^*(N),
    \psi_{\alpha_1}(z_1)\ldots\psi_{\alpha_N}(z_N)
    \Omega_0\rangle_l^{g=1} =\nonumber}\\
    & &\prod_{i<j}\left(\frac{
    \Theta[{1/2\atop 1/2}](\frac{z_i-z_j}{L_x}|\tau)}{
    \del_z\Theta[{1/2\atop 1/2}](0|\tau)}\right)^{2\alpha_i\alpha_j}
    \Theta\!\left[{l/\alpha_0^2\atop 0}\right]\!\left(\left.
    \frac{Z\alpha_0^2}{L_x}\right|\tau\alpha_0^2\right)\,,
  \end{eqnarray}
where $Z = \sum_{i=1}^{n}\alpha_i z_i/\alpha_0$ is the center of charge
coordinate. Here we have introduced the $\Theta$-functions with
characteristic,
  $$
    \Theta\left[{a\atop b}\right](z|\tau) =
    \sum_{n\in\BZ}e^{2i\pi\tau\frac{(n+a)^2}{2} + 2i\pi(n+a)(z+b)}\,.
  $$
In the case of the Laughlin wave functions we get the well known result
  \begin{eqnarray}\label{eq:myvevtorus}\lefteqn{%
    \langle\Omega_{-\alpha_0}^*(N),\prod_{i=1}^N\psi_{\sqrt{2p+1}}(z_i)
    \Omega_0\rangle_l^{g=1} =\nonumber}\\
    & &\prod_{i<j}\left(\frac{
    \Theta[{1/2\atop 1/2}](\frac{z_i-z_j}{L_x}|\tau)}{
    \Theta'[{1/2\atop 1/2}](0|\tau)}\right)^{2p+1}
    \Theta\!\left[{l/(2p+1)\atop 0}\right]\!\left(\left.\frac{(2p+1)Z}{L_x}
    \right|(2p+1)\tau\right)\,,
  \end{eqnarray}
which now is $(2p+1)$-fold degenerated. This degeneracy stems from the
possibility of an additional quantum number carried by the ground state with
background charge on the torus and follows from the properties of the
representation of the braid group on the torus.
We denote the functions (\ref{eq:myvevtorus}) by $\psi_{p,l}(z_1,\ldots,z_n)$.
They form a $(2p+1)$-dimensional space closed under the action of the
magnetic translations $S_a$ and $T_b$. For the elementary translations by
Steps $a=L_x/(2p+1)$ and $b=L_y/(2p+1)$ one has
  $$
    S_a\psi_{p,l} = e^{\pi i\frac{l}{2p+1}}\psi_{p,l}\,,\quad\quad
    T_b\psi_{p,l} = \psi_{p,l+1}\,.
  $$
Moreover, $\psi_{p,l}$ transforms covariantly under the exchange of $L_x$ and
$L_y$, i.e.\ $\tau\rightarrow-1/\tau$, since
  \begin{eqnarray*}\lefteqn{%
    \Theta\!\left[{l/(2p+1)\atop 0}\right]\!\left(\left.\frac{z\sqrt{2p+1}}
    {\tau}\right|-\frac{(2p+1)}{\tau}\right) =}\\
    & & e^{i\pi z^2/\tau}\sqrt{\frac{\tau}{i(2p+1)}}\sum_{l'=1}^{2p+1}
    e^{-2\pi i\frac{ll'}{2p+1}}
    \Theta\!\left[{l'/(2p+1)\atop 0}\right]\!\left(\left.z\sqrt{2p+1}
    \right|-(2p+1)\tau\right)\,.
  \end{eqnarray*}
Altough here we have discussed only the case of the Laughlin wave functions,
there
exist generalizations of this approach to the Laughlin type wave functions
(\ref{eq:LT}) using rational non integer compactification radii.
%
%%< CFT -> PHASE TRANSITIONS >%%%%%%%%%%%%%%%%%%%%%%%%%%%%%%%%%%%%%%%%%%%%
%
  \par
  \mysection{From Conformal Field Theory to Phase Transitions}
  \pn
Since the seminal work of A.A.~Belavin, A.M.~Polyakov and A.B.~Zamolodchikov
in 1984 \cite{BPZ84}, we have a deep theoretical understanding of the
universality classes of second order phase transitions of two-dimensional
statistical systems. Such a phase transition is related to a CFT, since
scaling invariance at the critical point implies full conformal invariance
of the statistical field theory in all known cases.
  \par
But nature is much richer, and there are two-dimensional systems with a
completely different phase transition behaviour -- such as the QHE. Here
the {\em phases itself\/} are described by chiral $c=1$ RCFTs. Therefore,
the transition between two phases must map {\em different\/} RCFTs into
each other. Moreover, the QHE phases have a very high symmetry due to the
incompressibility of the states. While phase transitions usually also show up
a very high symmetry, one cannot expect to keep the full symmetry of
non-singular area preserving diffeomorphisms, since the size of the system
must change (see section two). On the other hand, we might well expect to
have conformal invariance at the transition point, since we again have
scaling invariance.
  \par
  \subsection{The Modular Group in the Game}
  \pn
The phase diagrams of the QHE and similarly that of high-$T_c$
superconductivity
have been studied in much detail \cite{Kie91,Lut93,HLR93}. The mainpoint is
the assumption of an infinite discrete symmetry group acting on the parameter
space. This is nothing strange and, in fact, an old idea which lead to the
discovery of $S$-duality by J.~Cardy and E.~Rabinovici \cite{CaRa81}.
  \par
The most prominent infinite discrete group is the modular group
$\Gamma = {\rm PSL}(2,\BZ)$ which is the free span of $S = {0\ -1\choose
1\ \phantom{-}0}$ and $T = {1\ 1\choose 0\ 1}$ with the relations
$S^2 = (ST)^3 = \BI$. It operates on the upper half complex plane $\BH$ by
$S:\tau\mapsto-1/\tau$ and $T:\tau\mapsto\tau+1$ which we may extend to
include the real line. This group, or certain subgroups as the main
congruence subgroups $\Gamma(N) = \{{a\ b\choose c\ d}\in\Gamma\mid
a\equiv d\equiv 1\,{\rm mod}\,N, b\equiv c\equiv 0\,{\rm mod}\,N\}$,
presumably govern the phase diagram structure of many models in both,
condensed matter physics as well as string theories. Common to all these
models is the existence of an infinite number of {\em different\/} states
into which the model may condense in dependency of the parameters.
  \par
As we have seen above, the quantization of the magnetic flux motivates the
following discrete operations, which map QHE states into each other:
  \be
    \begin{array}{rcll}
      \nu&\mapsto& \nu/(2\nu + 1) & \ \ \ {\rm attaching\ two\ }\uparrow
      {\rm-flux\ quanta,}\\
      \nu&\mapsto& \nu/(2\nu - 1) & \ \ \ {\rm attaching\ two\ }\downarrow
      {\rm-flux\ quanta,}\\
      \nu&\mapsto& \nu + 1        & \ \ \ {\rm adding\ a\ further\ Landau\
      level,}\\
      \nu&\mapsto& 1 - \nu        & \ \ \ {\rm particle hole duality.}
    \end{array}
  \ee
The first three transformations generate the subgroup $\Gamma_T(2)$
of the modular group $\Gamma={\rm PSL}(2,\BZ)$, which is spanned by
$ST^{-2}S$ and $T$. Every real filling factor $\nu\in\BR$ can be arbitrarily
well approximated by an infinite continued fraction expansion generated by
words $\ldots U_{p_3}^{n_3}U_{p_2}^{n_2}U_{p_1}^{n_1}\in\Gamma_{T}(2)$,
where $U_{p}^{n} = (ST^{-2}S)^pT^n = {\phantom{p}1\ \phantom{2p}n\phantom{+1}
\choose 2p\ 2pn+1}$ and $n_i,p_i\neq 0\ \forall i>0$. Rational fillings are
given by finite continued fractions
  \begin{eqnarray}
    \nu &=& [n_1,2p_1,n_2,2p_2,n3,2p_3,\ldots,n_k,2p_k,n_{k+1}]\nonumber\\
        &=& n_{k+1} + \frac{1}{{\ds 2p_k +
                       \frac{1}{{\ds\ddots +
                        \frac{1}{{\ds n_2 +
                         \frac{1}{{\ds 2p_1 + \frac{1}{n_1}
                      }} }} }} }}\,,
  \end{eqnarray}
which can directly be interpreted in our graphical expansion of section two
as
the conductivity of a network of successive loops\footnote{This is always the
leading graph to a given loop order.}. As an example see (\ref{eq:sxy2}) as
the second order graph with $\nu = [n,2p,m,2q]$.
  \par
Given a finite word $U=\prod_{i=1}^k U_{p_i}^{n_i}\in\Gamma_T(2)$ we define
its length as $\ell(U) = \sum_{i=1}^k (n_i+2p_i)$ and its order as
${\cal O}(U) = k$, if $p_k\neq 0$, otherwise ${\cal O}(U) = k-1$. This means
that we consider fillings which are equivalent modulo 1 as of the same order.
Since $U_q^0U_p^n=U_{p+q}^n$ and $U_q^mU_0^n=U_q^{n+m}$, we assume that all
words are given in the form of minimal order and length.
  \par
The difference between this phase diagram structure and the one e.g.\ of
the oblique confinement phases in $\BZ_p$ lattice gauge theory is that
different phases correspond to in general different boundary CFTs.
  \par
Therefore, we are looking for a class of RCFTs which might be able to
furnish the proposed phase transitions between different QHE states.
Since both states, which we want to connect, are described by chiral RCFTs
with central charge $c=1$ but different boundary conditions given by their
compactification radii $R_1$ and $R_2$, the operator content of their
bulk CFTs will be different. Moreover, since they both have the same central
charge, they cannot be connected by a renormalization group flow via unitary
theories which always decreases the central charge. Nonetheless, they might
be connected by {\em non-unitary\/} CFTs as long as the {\em effective\/}
central charge remains constant. In non-unitary theories we have to
distinguish between the vacuum with its su(1,1)-invariance, and the state
of lowest energy, i.e.\ lowest $L_0$ eigen value $h_{{\it min}}<0$.
The effective central charge is then defined as $c_{{\it eff}} = c -
24h_{{\it min}} \geq 0$. Since we are looking for rational theories, it is
clear that they must posess an extended chiral symmetry algebra\footnote{
Let $k$ denote the number of bosonic, $l$ the number of fermionic generators
spanning the chiral symmetry algebra ${\cal W}$ of a RCFT. Then we strictly
have $0 < c_{{\it eff}} < k+l/2$ \cite{EFHHNV92}.}.
  \par
The RCFTs with $c_{{\it eff}} = 1$ have been completely classified
\cite{Gin88,DVV88,Kir89,Flo93,Flo94}. Besides the well known theories
with $c = 1$, which are the Gaussian models mentioned above\footnote{
There are also the $\BZ_2$-orbifolds with partition function
$(Z[2R^2] + 2Z[4] - Z[1])/2$ and the three exceptional solutions
$(2Z[9] + Z[4] - Z[1])/2$, $(Z[16] + Z[9] + Z[4] - Z[1])/2$ and
$(Z[25] + Z[9] + Z[4] - Z[1])/2$.}, there exist several classes of
non-unitary RCFTs:
  \par
The ``bosonic'' ones are the simplest, they have central
charge $c = 1 - 24pq$, maximal extended chiral symmetry algebra
${\cal W}(2,3pq)$, and partition function
  \be\label{eq:zbos}
    Z_{{\it bos}}[p/q,p'/q'] = {\ts\frac{1}{2}}(Z[p/q] + Z[p'/q'])\,,
  \ee
where we must have $p'q' - pq = 1$. Note, that this means that there exists
a RCFT with partition function (\ref{eq:zbos}) if and only if the two
parameters correspond to an element of the modular group, since we may
rewrite the condition as
  \be\label{eq:bosdet}
    {\rm det}\left(\begin{array}{cc} p' & p \\ q & q' \end{array}\right)=1\,.
  \ee
We have duality in both arguments, i.e.\
$Z_{{\it bos}}[p/q,p'/q'] = Z_{{\it bos}}[q/p,p'/q'] =
Z_{{\it bos}}[q/p,q'/p'] = Z_{{\it bos}}[p/q,q'/p']$.
It will soon become clear that this class of RCFTs generates the phase
transitions for a $\Gamma(1)$-phase diagram, i.e.\ where the infinite
discrete group acting on the parameter space is the full modular group.
  \par
The second class are the ``fermionic'' theories with central charge
$c = 1 - 12pq$, maximal extended chiral symmetry algebra
${\cal W}(2,\frac{3}{2}pq)$, and partition function
  \begin{eqnarray}\label{eq:zferm}
    Z_{{\it ferm}}[p/q,p'/q'] &=& {\ts\frac{1}{2}}
    (Z[2p/q] + Z[2p'/q'] + Z[p/2q] + Z[p'/2q']) \\
    &=& Z_{{\it bos}}[2p/q,2p'/q']
    + Z_{{\it bos}}[p/2q,p'/2q']\,,\nonumber
  \end{eqnarray}
where we must now have $p'q' - pq = 2$. Thus, the fermionic theories will
generate a phase diagram structure governed by $\Gamma(2)$ which is generated
by $T^2$ and $ST^2S$. Our condition takes the form
  \be\label{eq:fermdet}
    {\rm det}\left(\begin{array}{cc} p' & p \\ q & q' \end{array}\right)=2\,.
  \ee
It is worthwile, to emphasize a little pecularity: Matrices with determinant
two do not belong to $\Gamma(2)$. A nice property of $\Gamma(2)$ acting on
rational numbers is to preserve parity in both, numerator and denominator,
separately. Our condition together with the fact that $p,q$ must both be odd
to yield a fermionic theory selects the equivalence class of completely odd
rational numbers. Choosing an arbitrary matrix of $\Gamma(2)$ we can select
the three possible equivalence classes by multiplying it with one of the
matrices ${2\ 0\choose 1\ 1}$, ${1\ 1\choose 0\ 2}$ or
${1\ -1\choose 1\ \phantom{-}1}$ to get $(0/1)=0$, $(1/0)=\infty$ or $(1/1)=1$
respectively. Note that these three matrices all have determinant two.
  \par
These both sets of RCFTs have $\BZ_2$-orbifolds and also $N=1$ supersymmetric
extensions which will not be important to us in the scope of this paper.
Nonetheless, they may well play a r\^ole in other condensed matter systems
such as high-$T_c$ superconductivity where an appearence of $N=1$
supersymmetry has been conjectured.
  \par
  \subsection{Transitions between chiral RCFTs}
  \pn
In this section we will make use of a detailed knowledge of the representation
theory of the non-unitary theories mentioned above. The interested reader is
refered to \cite{Flo93,Flo94}. Since we want to describe aspects of the QHE,
we concentrate in the following mainly on the fermionic theories defined in
(\ref{eq:zferm}) and (\ref{eq:fermdet}).
  \par
The partition function (\ref{eq:zferm}) is nothing else than the sum of two
partition functions of fermionic $c=1$ RCFTs, $Z_{{\it ferm}}[p/q,p'/q'] =
(Z_{{\it bos}}[2p/q,p/2q] + Z_{{\it bos}}[2p'/q,p'/2q'])/2$, here expressed
in terms of (\ref{eq:zbos}). But what does it mean to sum partition functions?
The direct sum of (the underlying algebras of) different CFTs yields the
tensor product of their Hilbert spaces und thus the product of the partition
functions. The only meaning of the sum of partition functions can be that one
CFT lives on two disjunct boundaries. In fact, from the form of the partition
function we learn that the theory consists of two sectors each one belonging
to the specific periodicity conditions of one of the two compactification
radii, expressed in the statistics parameters $\theta_1/\pi = p'/q'$ and
$\theta_2/\pi = p/q$. But a closer look to the representation theory of
the maximal extended chiral symmetry algebra ${\cal W}(2,
{\ts\frac{3}{2}}p'q')$ shows that the decomposition of the Hilbert space
into a direct sum of irreducible lowest-weight representations does not
respect this sector structure.
  \par
To be more specific, the characters of the vacuum representation and of the
representation on the lowest-weight state $\mid h_{{\it min}}\rangle$ involve
modular forms from both sectors of the partition function decomposed according
(\ref{eq:zdecomp}),
  \be
    \chi_{{\it vac}}(\tau) = \sum_{n\in\BZ_+}\chi_{h_{n,n}}^{{\it Vir}}(\tau)
      = \frac{1}{2\eta(\tau)}\left(\Theta_{0,p'q'/2}(\tau) -
      \Theta_{0,pq/2}(\tau)\right)\,,
  \ee
and for $\chi_{h_{{\it min}}}(\tau)$ one has to replace the difference by the
sum. Here the conformal weights are denoted as usual by
$h_{r,s} = \frac{1}{4}\left((r\alpha_-+s\alpha_+)^2-(\alpha_-+\alpha_+)^2
\right)$, where $\alpha_{\pm} = \alpha_0\pm\sqrt{1+\alpha_0^2}$ and in our
case $\alpha_0 = \sqrt{p'q'/2}$.
  \par
The fact that some characters involve modular forms of different moduli has
highly non trivial consequences. Firstly, this means that the theory somehow
twists the different boundary conditions, such that it cannot be completely
decomposed into two disjunct parts with ``homogenous'' boundary conditions.
Therefore, the $S$-matrix, which describes the modular transformations
of the characters under $\tau\mapsto -1/\tau$, does not have block structure.
Secondly, the fusion rules of the RCFT, which can be obtained from the
the $S$-matrix via the Verlinde formula, have the property that fusing
two ${\cal W}$-conformal families from one sector can yield
${\cal W}$-conformal families of the other sector on the right hand side.
Exactly this property will enable us to shift from one QHE state to another.
  \par
Before we demonstrate how this works, we have to make one essential remark:
QHE states are described by {\em chiral} RCFTs, phase transitions by some
tensor product of a left- and right-chiral CFT to one, whose vertex operators
are all local and whose correlators are all well defined single valued
functions. If we look at the chiral part of one of our non-unitary RCFTs,
we find that it is isomorphic to a direct sum of two chiral $c=1$ RCFTs (if
we carefully work with effective central charge and effective lowest-weights).
Thus, we can recover the QHE states by appropriate {\em chiral projections\/}
of our non-unitary RCFT.
  \par
We start with the best understood states, the Laughlin states to
filling factors $\nu = 1/(2p+1)$. We rewrite the Laughlin wave functions as
correlators of local chiral vertex operators of a fermionic non-unitary
theory showing explicitely how we get the chiral QHE state as chiral
projection. Let us denote the full vertex operators by
$V_{(k,l\mid m,n)}(z,\bar z) \equiv\psi_{\alpha_{k,l}}(z)\otimes
\psi_{\alpha_{m,n}}(\bar z)$. We consider the non-unitary theory with
central charge $c = 1 - 24\frac{2p+1}{2}$ and chiral symmetry algebra
${\cal W}(2,\frac{3}{2}(2p+1))$. The diagonal partition function is
characterised by the matrix
${\phantom{2p}1\phantom{+}\ \phantom{p}1\phantom{2+}\choose 2p+3\ 2p+1}$.
Looking at the irreducible representation to $\mid h_{{\it min}}\rangle$ we
find a vertex operator $V_{(0,0\mid 1,1)}$ of conformal weight
$(h,\bar h)=(-(2p+1)/2,0)$. Its $N$-point correlator on the sphere is
  \begin{equation}\label{eq:myvev}
    \langle\Omega_{-\sqrt{(2p+1)/2}}^*(N),\prod_{i=1}^NV_{(0,0|1,1)}(z_i,
    \bar z_i)\Omega_0\rangle =
    \prod_{i<j}(z_i-z_j)^{2p+1}\exp\left(-\frac{1}{2}\sum_i|z_i|^2\right)\,,
  \end{equation}
hence identical with (\ref{eq:myvev0}). Actually, as a rigorous fact, the
chiral sectors of our non-unitary theories are indistinguishable from a
disjunct union of two chiral $c=1$ CFTs.
  \pano
Let us now assume that the external magnetic field is slowly increased.
Our present QHE state is realized by a uniform distribution of magnetic
flux and electrons such that each electron has $2p$ flux quanta attached to
it. Increasing the magnetic field inserts additional flux quanta which at the
beginning are not bound to an electron. We may now imagine that there will
be an amount of additional flux quanta such that a reconfiguration of the
system becomes possible in which again all flux quanta are somehow bound to
electrons. Certainly, there is at least the possibility to add $2N$ flux
quanta such that each electron could carry $2(p+1)$ at all. Note, that all
possible QHE phases described above are given by a certain partitioning of
uniform composite fermion subbands, i.e.\ electrons having all the
same number of attached flux quanta within the same subband, such that
{\em all\/} available flux quanta are eaten up that way.
  \pano
The reason for this is that the wave functions (\ref{eq:LT}) will be single
valued only, if there are no ``free'' flux quanta around. This explains, why
the filling factors are given by continued fractions, since the latter exactly
implement all electron density partitionings of the required kind.
  \pano
A closer look to the spectrum of our non-unitary theory in question shows
that there is a good candidate vertex operator for describing a single
flux quantum. While the composite fermions are described by
$V_{(0,0|1,1)}(z,\bar z)$, the flux quantum is realized by the operator
$V_{(\frac{2p}{2p+1},\frac{2p}{2p+1}|\frac{2p}{2p+1},\frac{2p}{2p+1})}
(w,\bar w)$, whose conformal dimension is $(h,\bar h)=(\frac{4p+1}{4p+2},
\frac{4p+1}{4p+2})$. If we insert $M$ such flux quanta into the correlator
(\ref{eq:myvev}), we obtain
  \begin{eqnarray}\label{eq:anyon}\lefteqn{%
    \langle\Omega_{-\sqrt{(2p+1)/2}}^*(N,M),\prod_{j=1}^M
    V_{(\frac{2p}{2p+1},\frac{2p}{2p+1}|\frac{2p}{2p+1},\frac{2p}{2p+1})}(w_j,
    \bar w_j)\prod_{i=1}^NV_{(0,0|1,1)}(z_i,\bar z_i)\Omega_0\rangle=}
    \nonumber\\
    & & \prod_{j<j'}|w_i-w_{j'}|^{1/(2p+1)}\prod_{i,j}(z_i-w_j)
    \prod_{i<i'}(z_i-z_{i'})^{2p+1}    %\\
    % & &\times\exp
    e^{-\frac{1}{2}\sum_i|z_i|^2-
    \frac{1}{2(2p+1)}\sum_j|w_i|^2}\,. %\nonumber
  \end{eqnarray}
Indeed, the flux quanta have the fractional statistics parameter $\theta/\pi
= 1/(2p+1)$, and the fractional charge $-e/(2p+1)$. Thus, they behave as
anyons \cite{Wil82,MR90}. In this way, we reproduce the basic excitations
of the Laughlin wave functions. Of cource, the anti-holomorphic part
$\prod_{j<j'}(\bar w_j - \bar w_{j'})^{1/2(2p+1)}$ drops out in the chiral
projection but cannot be avoided due to mathematical consistency.
  \pn
The main idea now is the following: If we read ``attaching of flux quanta''
literally, we must let approach the coordinates of the flux quanta to the
ones of the particles, $w_i\rightarrow z_i$. For simplicity let us first
consider the case $M=N$. Then we can attach one flux quantum to each particle.
We now insert the operator product expansion (OPE) of
$V_{(0,0\mid 1,1)}(z,\bar z)
V_{(\frac{2p}{2p+1},\frac{2p}{2p+1}|\frac{2p}{2p+1},\frac{2p}{2p+1})}
(w,\bar w)$ which is valid for $\mid z - w\mid\ll 1$. The OPE has the general
form
  \begin{eqnarray}\lefteqn{%
    V_{(\alpha\mid\beta)}(z,\bar z)V_{(\gamma\mid\delta)}(w,\bar w)
    =\nonumber}\\
    & &\sum_{\zeta,\eta}(z-w)^{h(\zeta)-h(\alpha)-h(\gamma)}
    (\bar z - \bar w)^{\bar h(\eta)-\bar h(\beta)-\bar h(\delta)}
    C_{\alpha\gamma}^{\zeta} C_{\beta\delta}^{\eta}
    \Phi_{(\zeta\mid\eta)}(w,\bar w)\,,
  \end{eqnarray}
where $\Phi_{(\zeta\mid\eta)}$ denotes a generic field $f(\del\phi,\del^2\phi
\ldots)V_{(\zeta\mid\eta)}$. The fusion rules of
our RCFT tell us which ${\cal W}$-conformal families will contribute to the
right hand side of the OPE. Since we want to take the limit $w\rightarrow z$,
we may restrict ourselfs to the term of leading order. Thus, with ``attaching
flux quanta'' we mean the fusion product, i.e.\ the projection of the OPE
to its leading order in the limit $w\rightarrow z$.
  \par
If we do this in (\ref{eq:anyon}) with $M=N$, we see that the right-chiral
part of $V_{(0,0|1,1)}(z,\bar z)$ acts as identity. Thus, inserting the OPE
we obtain the left-chiral part tensorized with
$\psi_{\alpha_{\frac{2p}{2p+1},\frac{2p}{2p+1}}}(\bar z)$. Obviously, this
has {\em no\/} purely chiral projection! But let us repeat this procedure
with a second set of $N$ flux quanta, hence attaching {\em two\/} flux quanta.
Studying the fusion rules of our RCFT \cite{Flo93}, we find the surprising
result that the leading right-chiral part again is the identity
$\psi_{\alpha_{1,1}}(\bar z)$, while in the left-chiral part the identity does
not appear and the leading vertex operator
$\psi_{\alpha_{\frac{2p+1}{2p+3},-\frac{2p+1}{2p+3}}}(z)$
belongs to the second sector of our RCFT! Thus, by chiral projection, we
moved from one $c=1$ CFT to another.
  \par
This shows, that attaching an {\em even\/} number of flux quanta, described
by fusion product, changes the periodicity conditions and the statistics
of the state from $1/(2p+1)$ to $1/(2p+3)$. In symbolical notation of fusion
we have for the left-chiral part
  \be\label{eq:fusel}
    \left(\left[\phi_{0,0}\right]\star
    \left[\phi_{\frac{2p}{2p+1},\frac{2p}{2p+1}}\right]\right)\star
    \left[\phi_{\frac{2p}{2p+1},\frac{2p}{2p+1}}\right] =
    \left[\phi_{\frac{2p+1}{2p+3},-\frac{2p+1}{2p+3}}\right] + \ldots
  \ee
to leading order. Up to this moment, we arrived at a $c=1$ CFT with
compactification radius $R^2 = 1/(2p+3)$, but with some excited state.
Nonetheless, the description of the phase transition is not yet complete,
since we still have to change the size of the system, such that the total
magnetic flux density remains constant \cite{Jai89a,Jai89b}. This dissipation
of the system (since we must decrease the electron density) will cost energy
to compensate the external pressure of the magnetic field,
and therefore, cool down the QHE state. Once we have changed the statistics by
the fusion product, the system cannot cool down to the old ground state, it is
forced to find a stable (hence chiral) state in the $c=1$ CFT with
compactification $R^2=1/(2p+3)$. This state is the Laughlin wave function
with $\nu = 1/(2p+3)$. The amount of energy will be proportional to the
ratio of the filling factors, $\nu_{{\it new}}/\nu_{{\it old}} = \nu^*/\nu =
(2p+1)/(2p+3)$. This is exactly the excitation energy of our leading term,
since $h(\alpha_{\frac{2p+1}{2p+3},-\frac{2p+1}{2p+3}}) = -(2p+1)/(2p+3)$.
  \par
Now we could start the game again since, after renormalization of the size
of the system, our starting point will be the non-unitary RCFT with
$c = 1 - 12(2p+3)$ and chiral symmetry algebra $\w(2,\frac{3}{2}(2p+3))$.
We will see, that the Laughlin states are characterized by the property
that the filling factor is equal to the statistics of the basic anyonic
excitations, $\nu=\theta/\pi$. More generally, we can form sequences of the
following kind. Let ${\cal C}[p/q]$ denote a chiral CFT with compactification
$R^2=p/q$, and ${\cal W}[p/q,r/s]$ a non-unitary fermionic RCFT with
$c_{{\it eff}} = 1$. Further let $\pi$ denote (left-) chiral projection.
Then we have sequences
  \be\label{eq:flow}
    {\cal C}[p_1/q_1] \stackrel{\pi^{-1}}{\longrightarrow}
    {\cal W}[p_1/q_1,p_2/q_2] \stackrel{\pi}{\longrightarrow}
    {\cal C}[p_2/q_2] \stackrel{\pi^{-1}}{\longrightarrow}
    {\cal W}[p_2/q_2,p_3/q_3] \stackrel{\pi}{\longrightarrow}
    \ldots\,,
  \ee
where according to general arguments on the renormalization group flow
of CFTs we must have a decreasing central charge in direction of
the flow, i.e.\ $c_1 = 1 - 12p_1q_1 > c_2 = 1 - 12p_2q_2 > \ldots$. This
in turn means that the product of $p_iq_i$ has to increase along the
sequence.
  \par
In this way we can obtain other filling factors with odd numerators and
denominaters. Let us start with the filling $\nu = p/q'$, given by a certain
continued fraction. Let us choose a $c=1$ theory with suitable
compactification radius, i.e.\ $R^2 = p'/q'$, which at the same time selects
the statistics parameter of anyonic excitations of our QHE state to be
$\theta/\pi = p'/q'$. Now, if it is possible to find an positive integer $q$,
such that (\ref{eq:fermdet}) can be fulfilled, we found a possible
QHE state with filling factor $\nu^*=p'/q$ given by a $c=1$ CFT with
$R^2=p/q$. Note, that the condition (\ref{eq:fermdet}) is non trivial and
restricts the possible phase transitions between QHE states.
  \par
For every admissible filling $\nu=p/q'$ where is a infinite set of fillings
$\nu^*_i=p'_i/q_i$, such that (\ref{eq:fermdet}) is fulfilled. For example,
$\nu=p/q'$ has ${p\ \phantom{q'}p'\phantom{+2}\choose q'\ p'q'+2}$ as
admissible
matrices allowing transitions to $\nu^*=p'/(p'q'+2)$ for $p'>0$ odd. On the
other hand, for every statistics parameter $\theta/\pi = p/q$ where is only
a finite (but never empty) set of statistical parameters $\theta^*/\pi =
p'/q'$ such that condition (\ref{eq:fermdet}) can be satisfied. Therefore,
a given statistics parameter can yield only a finite set of filling factors
and QHE states.
  \par
If $p,p'\neq 1$, the partition function (\ref{eq:zferm}) is no longer
diagonal. This affects the fusion rules by an automorphism \cite{Flo93}
and will in general change our leading order in (\ref{eq:fusel}). Moreover,
the non trivial factorization $p'q'$ or $pq$ guarantees that there are chiral
vertex operators of (half-) integer conformal weight above the ground state
vertex operator $\psi_{\alpha_{0,0}}$ such that we can have new chiral
projections with single valued correlators, i.e.\ wave functions of the more
general form (\ref{eq:LT}).
%
%%< PHASE TRANSITIONS -> FRACTALS >%%%%%%%%%%%%%%%%%%%%%%%%%%%%%%%%%%%%%%%
%
  \par
  \mysection{From Phase Transitions to Fractals}
  \pn
Let us return for a moment to our class of non-unitary RCFTs with
$c_{{\it eff}} = 1$ and condition (\ref{eq:fermdet}) for the fermionic case.
Our partition function depends on two parameters. We have studied the
moduli space of these theories in much detail, see \cite{Flo94} for the
bosonic case. The strange fact that we are not allowed to combine two
arbitrary rational $c=1$ CFTs, i.e.\ combine two Gaussian partition functions
$Z[x],Z[y]$ with $x,y$ rational, in order to get a new {\em rational\/}
theory, is intimately related to the modular group. As has been explained
above, we can obtain all admissible partition functions by the
orbit of $\Gamma$ (or a subgroup as $\Gamma(2)$ for the fermionic case) on
a suitable matrix, in our case ${1\ -1\choose 1\ \phantom{-}1}$.
  \par
The set of all admissible pairs $(x,y)\in(\BR_+)^2$ yields a multi fractal
of measure zero (actually a Cantor set) which nonetheless lies dense in the
plane\footnote{The proof of these facts can be found in \cite{Flo94} for the
bosonic case. It is easy to see that things remain true in the fermionic
case, since $\Gamma(2)\subset\Gamma$ is an infinite discrete subgroup of
index two.}. figure 1 shows a crude approximation of it, where we mapped all
pairs into the fundamental region $\{(x,y)\in(\BR_+)^2\mid 0\leq x\leq y\leq
1\}$ via duality of the partition function in each of its arguments. Due to
purely esthetic reasons we mirrored everything at the diagonal.
  \begin{figure}[hbt]
%%%%% FIGURES --- BEGIN %%%%%%%%%%%%%%%%%%%%%%%%%%%%%%%%%%%%%%%%%%%%%%%%%%%
%
  \def\epsfsize#1#2{1.5#1}
  %\begin{center}
  %\hbox{$\phantom{xxxxxxxxxxxxxxx}$\
  \hbox{\
% \epsfbox{c:\michael\ps\fra.ps}
  \epsfbox{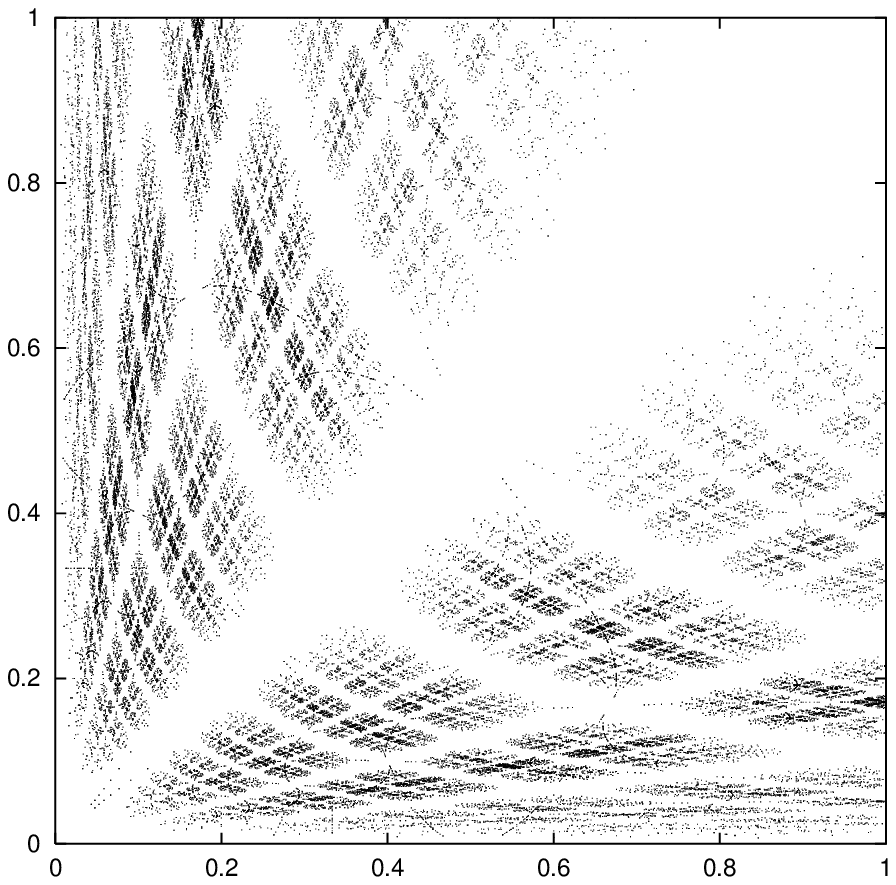}
  }
  %\end{center}
  \def\epsfsize#1#2{1.0#1}
%
%%%%% FIGURES --- END %%%%%%%%%%%%%%%%%%%%%%%%%%%%%%%%%%%%%%%%%%%%%%%%%%%%%
  \pn
  \begin{quotation}
  \noindent{\small {\sc Figure 1.} The moduli space of the fermionic
  $c_{{\it eff}} = 1$ theories with partition function
  $Z_{{\it ferm}}[p/q,p',q']$. Plotted are all admissible pairs
  $(x=p/q,y=p'/q')$ of completely odd rational numbers which fulfill
  $p'q'-pq = 2$. The plot is restricted to $(x,y)\in[0,1]\times[0,1]$ since
  points outside this region are identified via duality of the partition
  function. The plot is generated from matrices $A\cdot{1\ -1\choose
  1\ \phantom{-}1}$, $A\in\Gamma(2)$, with length $\ell(A)\leq 10$.
  }\end{quotation}
  \end{figure}
  \par
  \subsection{Particle-Hole Duality}
  \pn
At this point the interested reader may wonder, where the filling factors
with even numerators appear. Up to now, we implemented the following
operations on filling factors:
  \be
    \begin{array}{rcll}
      \nu&\mapsto& \nu/(2\nu + 1) & \ \ \ {\rm attaching\ two\ }\uparrow
      {\rm -flux\ quanta,}\\
      \nu&\mapsto& \nu + 2        & \ \ \ {\rm adding\ two\ further\ Landau\
      levels,}
    \end{array}
  \ee
which are generated by the main congruence subgroup $\Gamma(2)$ of the
modular group $\Gamma$, spanned by $ST^{-2}S$ and $T^2$. This subgroup
preserves the parity of numerator and denominator separately.
Our condition on the admissible partition functions chooses the odd-odd
parity, assured by multiplying a matrix $A\in\Gamma(2)$ with
${1\ -1\choose 1\ \phantom{-}1}$.
  \par
Firstly, if we change the direction of the external magnetic field, we
easily obtain a mapping between fractions of the form
  \be
    \nu\mapsto\nu/(2\nu - 1)\ \ \ {\rm attaching\ two\ }\downarrow
    {\rm -flux\ quanta,}
  \ee
where we just replace $ST^{-2}S$ by its inverse $ST^2S$. This does not
change the parity. But it changes the sign of the determinant
(\ref{eq:fermdet}). Thus, we have to exchange the columns of the matrix
which gives us a canceling sign. We see that in this way our ``flow'' between
QHE states according (\ref{eq:flow}) keeps its direction which never
decreases denominators, i.e.\ which never leads to ``less anyonic''
statistics.
  \par
But there is another mapping of QHE states, the so called particle-hole
duality which transforms filling fractions as
  \be
    \nu\mapsto1-\nu\ \ \ {\rm particle-hole\ duality.}
  \ee
This transformation is not contained in $\Gamma(2)$, even not in the whole
modular group. But it has a natural explanation within our scheme.
Our partition function $Z_{{\em ferm}}[p/q,p'/q']$ posesses duality in each
of its arguments. This means that the compactification and statistics $p/q$
have the same spectrum as $q/p$, and therefore, can generate the same QHE
states. The analogous holds for $p'/q'$. Duality corresponds to exchanging
the elements of the main- or off-diagonal of the matrix $A = {p\ p'\choose
q'\ q}$, which we denote by $D_{/}$ or $D_{\backslash}$ respective. Obviously
we have $D_{\backslash}D_{/} = D_{/}D_{\backslash} = D_{\times}$.
We see that $D_{\times}$ just exchanges the order of the compactification
radii, since $p/q < p'/q'$ implies $q'/p' < q/p$. This also changes the
direction of our ``flow'' between QHE states and it might happen that
a QHE state is mapped to one with less anyonic statistics. But this is
impossible from the sector structure of our theories, since the fusion
product always leads to vertex operators of more anyonic statistics.
Moreover, as a consequence, the renormalization group flow between the
non-unitary RCFTs with $c_{{\it eff}} = 1$ would not decrease the
central charge. The same happens, if we apply the transformation
$TS\not\in\Gamma(2)$ to $A$ which maps $\nu$ to $1 - 1/\nu$.
  \par
Combining both, we obtain what we want. In fact, the pair
$(\nu,\nu^*) = (p/q',p'/q)$ is mapped to $(1-\nu^*, 1-\nu) =
(1-p'/q,1-p/q')$, thus yielding the particle-hole duality. The funny point
is now that this operation changes the parity of the numerators! In this
way we obtain rational numbers with even numerator but still odd denominator.
  \par
We pause here to remark that in the frame of QHE the word hole does not
mean the same as in general semiconductor theory. Here a hole is realized
by quasiparticles consisting of an appropriate number of flux quanta. It is
well known \cite{MR90,Lau83,Sto92,PG87} that an inserted single flux
quantum behaves as a quasihole of charge $q=-e/(2p+1)$ and statistics
$\theta/\pi=1/(2p+1)$, see (\ref{eq:anyon}). So, $2p+1$ such quasiholes make
up a ``real'' fermionic hole of charge $q=-e$ and statistics $\theta/\pi = 1$.
But there is a slight difference. While real electrons have antiperiodic
boundary conditions, since they {\em have\/} the freedom to change their
spin relatively to the external magnetic field (although it is unlikely),
flux quanta have not. This we can incorporate by giving a hole fermionic,
but periodic boundary conditions. This is by no way unnatural. We only change
from the Neveu-Schwarz sector of a fermionic CFT to its Ramond sector,
and naturally we will get statistics with even numerators.
  \par
Therefore, depending on whether we live in Neveu-Schwarz or in Ramond sector,
our RCFT in question will describe a transition between two electron QHE
states or between two hole QHE states. The transitions by itself preserve
parity.
  \par
In this discussion we used the fact that we are not allowed to use duality
of the partition function without redefining the filling fraction such that
it remains invariant. Otherwise we could not avoid to get transitions in
the wrong direction. This has two consequences. Firstly, there will be no
trouble with even denominators after using particle-hole duality, which gives
us even numerator fillings. Secondly, we do not obtain the pure IQHE states
by this procedure, since we cannot use duality on the Laughlin states. But
we do not expect to obtain IQHE states (except $\nu=1=1/(2p-1)$ with $p=1$ and
its descendents), since they lack a real Chern-Simons interaction and
therefore, are not described by a Lagrangian of the form (\ref{eq:csll}).
  \par
  \subsection{Attractors and Fractals}
  \pn
Finally, we would like to explain the plateaux within our scheme. Let us
consider a sequence of type (\ref{eq:flow}) and its induced sequence of
matrices ${p_2\ p_1\choose q_1\ q_2}\longrightarrow
{p_3\ p_2\choose q_2\ q_3}\longrightarrow
{p_4\ p_3\choose q_3\ q_4}\longrightarrow\ldots$. As is well known, the
corresponding sequence of filling factors will converge to a real number,
if we assure that the length $\ell(A_i)$ does not decrease. But this is
exactly achieved by the direction of the renormalization group flow.
Let us consider a sequence, where even the order ${\cal O}(A_i)$ is strictly
increasing. Then the sequence of pairs $(\nu_i=p_{i+1}/q_i,\nu^*_i=
p_i/q_{i+1})$ converges to $(\nu_{\infty}=\nu^*_{\infty})$ where
$\nu_{\infty}$ lies somehow in the intervall $[\nu_1,\nu^*_1]$. Actually,
starting with a matrix $A_1$, all matrices $B=A\cdot A_1$ such that
${\cal O}(B)>{\cal O}(A_1)$ correspond to filling fractions inside this
intervall. Matrices with this property are denoted by $B\succ A_1$. It has
been shown in \cite{Flo94} that the set of all such matrices, i.e.\ of all
pairs of compactifications $(p_i/q_i,p_{i+1}/q_{i+1})$ plotted as in
figure 1 (but without identfying points via duality), is confined to a band
in the following way: Let $w^2$ be the product of the compactifications,
$w^2 = (p_i/q_i)(p_{i+1}/q_{i+1}) = \nu_i\nu^*_i$. Then the orbit of all
matrices $B = A\cdot A_1 \succ A_1$ is confined to a hyperbolic band of width
$\varepsilon=2w\Delta\nu + (\Delta\nu)^2$, where $\Delta\nu = \mid\nu_1 -
\nu^*_1\mid$, defined by the equation $x/y = \alpha$ with $\alpha\in
[w^2-\varepsilon,w^2+\varepsilon]$.
  \par
Let us choose $\nu_1$ and $A_1$ of order ${\cal O}(A_1) = 1$.
Then the orbit of
all matrices $B\succ A_1$ will intersect the diagonal in figure 1 at a point
$x$ near $w$. The diagonal corresponds to unitary fermionic $c=1$ CFTs, since
$Z_{{\it ferm}}[x,x] = Z[2x]+Z[x/2]$. This unitary theory is a never reached
fixpoint for a certain sequence of matrices $B\succ A_1$ with $\nu_{\infty}
= x$. Moreover, if we can truncate our $1/N$-expansion to first order,
all filling fractions corresponding to matrices $B\succ A_1$
will have the Hall conductivity $\sigma_{xy} = \nu_1$, since $\nu_1$ is the
first order continued fraction expansion for all $\nu(B)$, $B\succ A_1$.
We call the set of all points $(\nu(B),\nu^*(B))$ to $B\succ A_1$ the
attractor band of $(\nu(A_1),\nu^*(A_1))$.
  \par
Physically this means the following: Let us start from a QHE state with
filling fraction $\nu = [n_1,2p_1,n_2,2p_2,\ldots,2p_k]$ and let us first
assume that the Hall conductivity is given by a first order effect,
$\sigma_{xy} = [n_1,2p_1,n_2]$. If we increase the external magnetic field
very slowly, the filling fraction may change a bit. If the change of the
magnetic field is small enough, $\nu$ will just change by a small amount
coming from minor corrections in the highest orders of its continued fraction
expansion, eventually the order itself may change a bit. This
corresponds to a reconfiguration of the system and the Lagrangian
(\ref{eq:csll}) such that again all flux quanta are bounded to electrons.
Since the Hall conductivity is a first order effect, it does not change. Thus,
we are moving inside an attractor band of the kind described above.
  \par
If now the increment of the external magnetic field is strong enough such that
$\nu$ changes to a number expressible in a continued fraction expansion, whose
{\em first\/} orders are different, the Hall conductivity changes. We heve
then moved from one first order band to another. How can this happen, if our
flow of theories never decreases the order? Firstly, as we have seen
for the Laughlin states, there are transitions between fractions of the
{\em same\/} order, and in this case the first order changed,
such that the Hall conductivity changed too. Secondly, if we approach
a rational number with a short continued fraction by a sequence of
continued fractions of larger length, the system will certainly choose the
much simpler inner configuration as soon as it can, since this decreases the
number of states. Consideration of a sequence of longer and longer continued
fractions assumes at the same time, that the magnetic field is increased
slower and slower. But there is a limit, since we either insert at least one
flux quantum, or nothing changes. Thus, the general situation will be that
the increment of the magnetic field will affect lower orders too, since it
is done within a finite time.
  \par
figure 2 shows a crude approximation of the attractor bands in a double
logarithmic plot (where we inverted the hyperbolas to origin lines,
$x = 2R_1^2, y = 1/2R_2^2$, which
become parallel due to the logarithmic scales). We calculated the bands only
with matrices $A\cdot A_1$ with $\ell(A)\leq 10$, since otherwise the plot
would be overcrowded and neighbouring bands would be undistinguishable. One
sees, that the bands can have small overlaps which
are due to possible transitions which change the first order in the continued
fraction. In this figure we only show the bands to the experimentally
observed Hall conductivities of first order. There are other observed
Hall conductivities, e.g.\ the both hole QHE states $\frac{4}{11} = [1,2,1,2]$
and $\frac{4}{13} = [-2,2,-2,4]$, which are of higher order and have non
vanishing longitudinal conductivity. They presumably belong to Hall samples,
where the number $N$ of electrons (or holes) is small enough to allow second
order effects to contribute.
  \par
The attractor bands are defined relatively to a start {\em pair\/} of
fillings $(\nu_1,\nu_1^*)$. For a given $\nu_1$ we choose $\nu_1^*$ such
that the statistics parameters have the smallest possible denominators.
Usually, this implies the largest possible $\Delta\nu$ resulting in some
overlap of the bands.
  \begin{figure}[hbt]
%%%%% FIGURES --- BEGIN %%%%%%%%%%%%%%%%%%%%%%%%%%%%%%%%%%%%%%%%%%%%%%%%%%%
%
  \def\epsfsize#1#2{1.5#1}
  %\begin{center}
  %\hbox{$\phantom{xxxxxxxxxxxxxxx}$\
  \hbox{\
% \epsfbox{c:\michael\ps\fqhe0.ps}
  \epsfbox{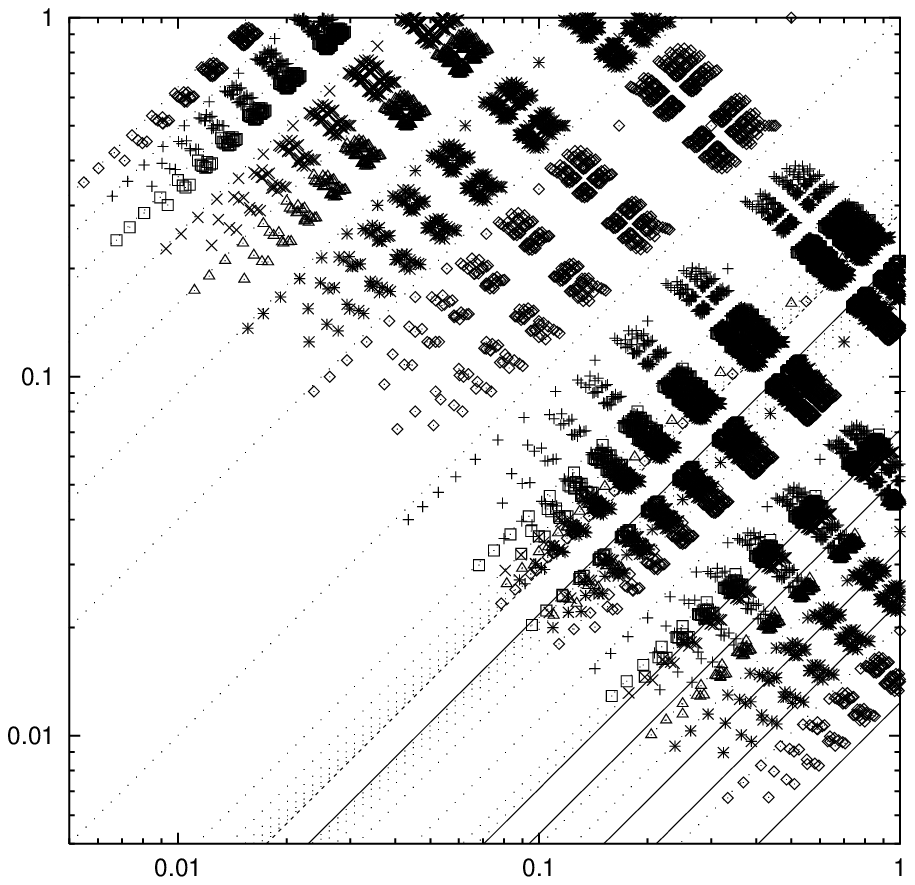}
  }
  %\end{center}
  \def\epsfsize#1#2{1.0#1}
%
%%%%% FIGURES --- END %%%%%%%%%%%%%%%%%%%%%%%%%%%%%%%%%%%%%%%%%%%%%%%%%%%%%
  \pn
  \begin{quotation}
  \noindent{\small {\sc Figure 2.} Attractor bands of all experimentally
  observed Hall plateaux of first order. We generated parallel lines by
  logarithmic scales in order to improve clarity of the plot. Plotted are
  all points $(\nu,\nu^*)$ to matrices $A'\cdot A$ with $\ell(A')\leq 10$.
  Here the start matrix $A$ is choosen as described in the text. From top
  left to bottom order the different fractions are
  $\nu\in\{8,7,6,5,4,3,2,1,\frac{2}{3},\frac{3}{5},\frac{4}{7},
  \frac{5}{9},\frac{6}{11},\frac{7}{13},\frac{6}{13},\frac{5}{11},
  \frac{4}{9},\frac{3}{7},\frac{2}{5},\frac{1}{3},\frac{2}{7},
  \frac{3}{11},\frac{2}{9},\frac{1}{5},\frac{1}{7}\}$, distinguished by
  the plot symbols.
  }\end{quotation}
  \end{figure}
%
%%< SUMMARY -> DISCUSSION >%%%%%%%%%%%%%%%%%%%%%%%%%%%%%%%%%%%%%%%%%%%%%%%
%
  \par
  \mysection{From Summary to Discussion}
  \pn
The aim of this paper was to introduce a new class of phase transitions
in two dimensions and, treated as an example, to explain the main features
of the QHE with this kind of transitions. Since this class is defined by
general considerations on RCFT, it can be applied to similar phenomena
of condensed matter physics which are essential two-dimensional. The main
assumption is a phase diagram whose topological structure is completely
determined by an infinite discrete group such as the modular group. In our
case the group in question is $\Gamma(2)$.
  \par
In a forthcoming work we will study high-$T_c$ superconductivity which is
supposed to have a very similar phase diagram as QHE. Due to a possible
decoupling of the electron charges from spin we expect that the phase
transitions may be described by $N=1$ supersymmetric extensions of the
RCFTs used in this work.
  \par
The logic of this paper works with two strategies.
  \par
Firstly, following a
work of J.~Fr\"ohlich and A.~Zee \cite{FZ91} showing that attaching flux
quanta is an in first order globel Chern-Simons interaction which couples
to the overall current of the electrons, we develop a graphical description
of Chern-Simons interactions in $1/N$-expansion, which is similar to the
Feynman graphs. These graphs give us a simple way to read off the
filling factor $\nu$. The macroscopic conductivity observables are obtained
by truncation to the maximal macroscopic contributiong order, i.e.\ for
$N\gg1$ only the first order. We argued that the graph may be viewed as
a classical conductor network, since the macroscopic observable currents
are subject to classical electrodynamics. Then the topological nature
of Chern-Simons-QED shows up in the way that there is no difference between
conductance and conductivity. The fact that there exists a first
order, i.e.\ that there is a Chern-Simons term in the Lagrangian which will
dominate the large scale physics, is related to the existence of impurities
at which the flux quanta are localized.
  \par
It depends crucialy on the experimental situation, whether and how many
fractional QHE states can be observed between the IQHE states. Typically,
the FQHE can only be observed at a tenth of the IQHE temperature range, it
needs a stonger magnetic field and a carefully choosen range of the impurity
density. But a further improvment of these presumptions seems not to lead to
a more refined plateau structure. Nearly all measured FQHE conductivities
are of first order, i.e.\ of the form
$\sigma_{xy} = \frac{n}{2pn\pm 1} + m$, $n\in\BN$, $m,p\in\BZ_+$. This
supports the theoretical prediction that the Chern-Simons interaction
is macroscopically of first order only.
  \par
Therefore, we interpret our graphs on one hand classicaly, and then only to
first order, and on the other hand as Feynman graphs of a Cherm-Simons-QED.
Since the graphs also encode a continued fraction expansion of the filling
fraction, we see that higher orders contribute smaller corrections to the
filling factor.
  \par
Secondly, using the equivalence of $(2+1)$-dimensional Chern-Simons theory
with chiral $(1+1)$-dimensional CFT, we construct transitions between QHE
states by non-unitary RCFTs with $c_{{\em eff}} = 1$, which connect two
different chiral $c=1$ CFTs. The structure of the moduli space of these
theories incorporates the subgroup $\Gamma(2)$ of the modular group which
is in agreement with theoretical results on the phase diagramm of the QHE.
Moreover, it predicts certain fixpoints and attractor bands which are related
to the observed plateau widths.
  \par
Every matrix $A$ defines a hyperbolic band in the moduli space
which contains all points to matrices $B\succ A$. The width of the band
is parametrised by $\Delta\nu = \mid\nu(A)-\nu^*(A)\mid$, the average
opening width of the hyperbolic band by $\nu(A)\cdot\nu^*(A)$. The pairs of
filling factors $(\nu(B),\nu^*(B))$ for all matrices $B\succ A$ yield
hyperbolic bands contained in the one of $A$. Mapping all matrices
$B = {p'\ p\choose q\ q'}\succ A$ by
  \begin{equation}\label{eq:newmap}
    Q:B = \left(\begin{array}{cc} p'& p\\
                                  q & q'\end{array}\right)\mapsto
    \left(\frac{p'}{q'},\frac{p}{q}\right)
  \end{equation}
into the plane fills a hyperbolic band of width $\varepsilon$,
  \be\label{eq:width}
    \varepsilon\sim 2\bar{\nu}\Delta\nu+(\Delta\nu)^2\sim 2\sqrt{\nu\cdot\nu^*
    }\mid\nu-\nu^*\mid + (\nu-\nu^*)^2\,.
  \ee
Note that in figure 1 the parts of the
hyperbolic bands with $x>1$ or $y>1$ are folded inside the plot and would
appear there as origin lines. In figure 3 we show just one single attractor
band, plotted in the way of figure 1.
  \begin{figure}[hbt]
%%%%% FIGURES --- BEGIN %%%%%%%%%%%%%%%%%%%%%%%%%%%%%%%%%%%%%%%%%%%%%%%%%%%
%
  \def\epsfsize#1#2{1.5#1}
  %\begin{center}
  %\hbox{$\phantom{xxxxxxxxxxxxxxx}$\
  \hbox{\
% \epsfbox{c:\michael\ps\attr.ps}
  \epsfbox{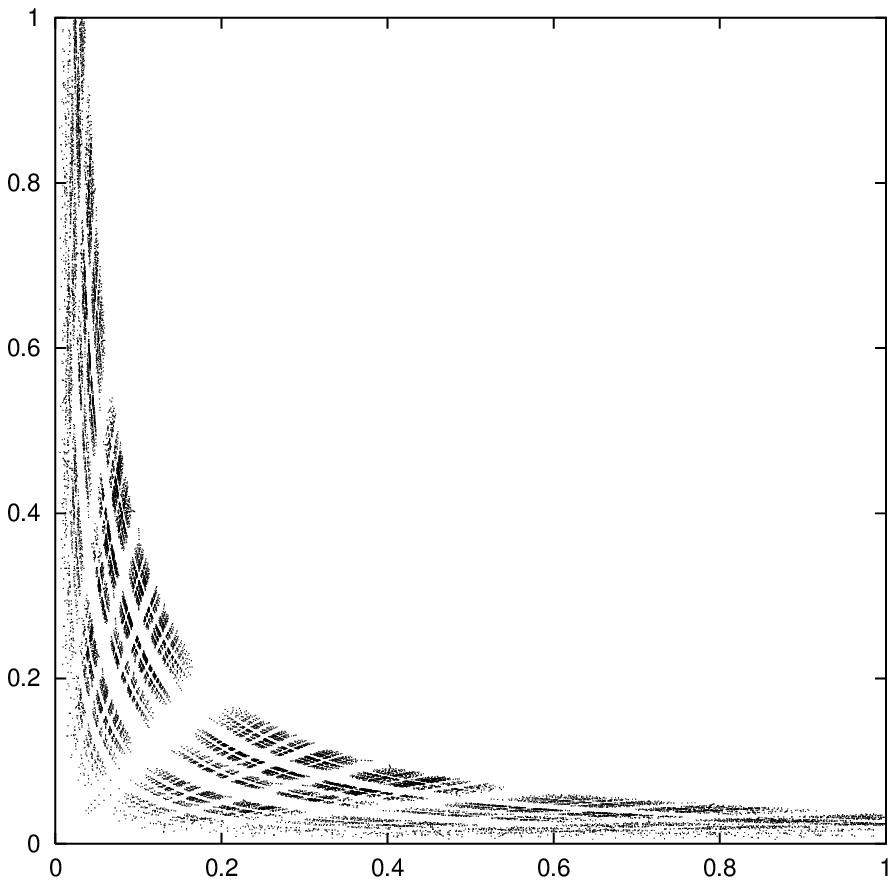}
  }
  %\end{center}
  \def\epsfsize#1#2{1.0#1}
%
%%%%% FIGURES --- END %%%%%%%%%%%%%%%%%%%%%%%%%%%%%%%%%%%%%%%%%%%%%%%%%%%%%
  \pn
  \begin{quotation}
  \noindent{\small {\sc Figure 3.} Attractor band to the matrix
  ${1\ 1\choose 5\ 7}$ which corresponds to all QHE state transitions whose
  new filling factors lie in $[1/7,1/5]$. Since nearly all such QHE states
  will have the Hall conductivity $\sigma_{xy} = 1/5$ (or eventually $1/7$)
  to first order, this defines a region of variation of the magnetic field,
  where the Hall conductivity remains constant. The mean slope of the
  broad lines is $\nu/\nu^*$ and $\nu^*/\nu$ respective. The dirty dust
  inside is an artefact of the used algorithm.
  }\end{quotation}
  \end{figure}
  \par
{}From our $1/N$ argumentation we conclude that (nearly) all observed
Hall conductivities $\sigma_{xy} = [n,2p,m]$ are given by first order
effects. The corresponding matrices $A\in\Gamma(2)$ with $(\nu=\sigma_{xy},
\nu^*)$ with $\Delta\nu$ maximal generate as germs of orbits $B = A'\cdot A$,
$A'\in\Gamma(2)$, hyperbolic bands of maximal width. These matrices $A$
correspond in general to transitions which change the first order of the
continued fraction expansions, since $\nu$ is supposed to be of first order
$\Delta\nu$ is choosen maximally. Thus, they define a range of variations of
the filling factor due to variation of the external magnetic field, where the
Hall conductivity remains constant.
  \par
Let the Hall sample be in an arbitrary generic state with an smeared out
filling factor $\bar{\nu}\pm\delta\nu$, since there are fluctuations
of thermal or quantum mechanical kind, or the external magnetic field
fluctuates. The system may no rearrange itself to a state of simpler
form, e.g.\ a state corresponding to a first order graph Lagrangian
(\ref{eq:csll}), which lies inside the attractor band to the state
$\bar{\nu}\pm\delta\nu$. In any case, the system will ``cool down'' to
the simplest possible state within the allowed $\nu$-range. The attractor
bands then define equivalence classes of QHE states, between which the
system can change without affecting the macroscopic observables. Since the
filling factor $\bar{\nu}$ is to first order a linear function of the
external magnetic field (similar to the {\em classical\/} Hall conductivity),
the attractor band widths directly correspond to plateau widths of the
quantum Hall conductivity. We have the surprising result that a macroscopic
observable is quantized while more quantum like parameters as the filling
factor vary more or less smoothly. The smoothness of $\nu$ is related to
the maximal possible order of the continued fraction. The latter in turn
depends directly on $N$ which is supposed to be very large.
  \par
Therefore, the plateau widths should be proportional to $\varepsilon$
according to (\ref{eq:width}). The experimental data do indeed support this.
The plateau width is the larger the smaller the numbers $n,p,m$ in the first
order expansion $\sigma_{xy} = [n,2p,m]$ are. Very small bands correspond to
very small plateaux which will naturally be difficult to observe, since they
are stable only in a very small range of the magnetic field. If the
fluctuations $\delta\nu$ are stronger then the width of the plateau, the
system will switch to a neighbouring plateau of larger width. This explains
the dominating of small numbers.
  \par
Finally, one can show that the attractor bands of the already experimentally
observed Hall conductivities cover the parameter plane almost completely.
The bands to the Laughlin states with $\sigma_{xy} = 1/(2p-1)$ have width
$4/(4p^2-1)^{3/2} + 4/(4p^2-1)^2$ centered around $\bar{\nu} =
(4p^2-1)^{-1/2}$ and thus, do not overlap. Between them there is space for
bands with $n>1$. In fact, we can cover the region ${\cal M} = \{(x,y)\in\BR^2
\,\mid\,x,y>0,\,xy<1\}$ with small numbers $n,p$. This region is shifted by
one along the diagonal under the operation of $T:m\mapsto m+1$. Therefore,
including particle-hole duality and changing of the direction of the
magnetic field, we can cover almost the whole region ${\cal M}$ with numbers
$p\leq 4, \mid n\mid \leq 8, p\mid n\mid < 8$, which correspond to the
experimentally observed Hall conductivities.
  \par
Figure 2 shows exactly this in a crude approximation. There seems to be some
space left between some bands. This is the case, if bands lie near
prominent ``forbidden'' zones corresponding to even denominator fillings,
which are difficult to approximate with matrices of small length (we plotted
only up to $\ell \leq 10$), see \cite{Flo94} for details. In fact, even
denominator fillings are not observed in ordinary Hall samples. In figure 2
we see such gaps for $\nu = 1/2$ and $\nu=1/4$. The region around $\nu=1$
seems to be rather empty, since approximation of the ``unitary'' line
$x=y$ is difficult with low order expansions.
In addition figure 2 shows with straight lines the position of that Hall
conductivities which should have the next stable plateaux, i.e.\ which
should show up in more precise experiments most likely. From top left to
bottom right they belong to the values
$\nu\in\{\frac{8}{15},\frac{7}{15},\frac{4}{15},\frac{3}{13},\frac{2}{11},
\frac{2}{13},\frac{1}{9}\}$. One sees that they further approch the
``forbidden'' zones $\nu = \frac{1}{2}$ and $\nu = \frac{1}{4}$.
  \par
If second order effects may contribute, the width of the attractor bands
is smaller, since $\Delta\nu$ has to be defined appropriately such that
the germ matrix corresponds to the maximal change of $\nu$ in the
second order. Then we naturally will have new gaps where Hall plateaux to
second order effects may fit in. It would be worthwile to study under which
experimental conditions the Hall sample supports second order QHE states,
i.e.\ under which circumstances the total number of electrons is low enough.
  \par
With the model of the FQHE proposed in this work, we have given an unified
view of several aspects, Chern-Simons theory, conformal field theory,
phase transitions, linked together by the modular group and its action
on a new class of non-unitary rational conformal field theories.
We used the FQHE as an example for a new class of phase transitions which
are related to these non-unitary RCFTs. Within this frame we are able to
explain the main features of the FQHE including the plateau widths and
the selection of observed plateaux.
  \par
The modular group once more showed up in theoretical physics, connecting
so different fields as arithmetic and fractal geometry, rational conformal
field theory, phase transitions in two dimensions, and -- most
fascinating -- experimentally observable real systems as the QHE.
  \pn
  \begin{center}
  {\bf Acknowledgement}
  \end{center}
  \pn
I am very grateful to C\'esar G\'omez and Eliezer Rabinovici for many
illuminating discussions and I would like to thank Jos\'e Gaite,
Werner Nahm, Germ\'an Sierra and Raimund Varnhagen for useful comments and
also Birgitt Federau for careful reading of the
manuscript. This work was supported by the Deutsche Forschungsgemeinschaft
and the ``Human Capital and Mobility'' program of the European Community.
  \newpage
%
%%< REFERENCES >%%%%%%%%%%%%%%%%%%%%%%%%%%%%%%%%%%%%%%%%%%%%%%%%%%%%%%%%%%
%
  

\begin{thebibliography}{99}
  {\footnotesize
\bibitem{AS85} {\sc J.F. Avron, R. Seiler}
   {\em Quantization of the Hall Conductance for General Multiparticle
   Schr\"odinger Hamiltonians},
   Phys. Rev. Lett. {\bf 54} (1985) 259-262,\\
   {\sc M. Klein, R. Seiler}
   {\em Power-Law Corrections to the Kubo Formula Vansih in Quantum
   Hall Systems},
   Commun. Math. Phys. {\bf 128} (1990) 141-160
\bibitem{Bak89} {\sc I. Bakas}
   {\em The Large $N$ Limit of Extended Conformal Symmetries},
   Phys. Lett. {\bf B228} (1989) 57
\bibitem{BBGS92} {\sc A.P. Balachandran, G. Bimonte, K.S. Gupta, A. Stern}
   {\em The Chern-Simons Source as a Conformal Family and its
   Vertex Operators},
   Int. Jour. Mod. Phys. {\bf A7} (1992) 5855-5876,
   {\em Conformal Edge Currents in Chern-Simons Theories},
   Int. J. Mod. Phys. {\bf A7} (1992) 4655-4670
\bibitem{BPZ84} {\sc A.A. Belavin, A.M. Polyakov, A.B. Zamolodchikov}
   {\em Infinite conformal symmetry in two-dimensional quantum field theory},
   Nucl. Phys. {\bf B241} (1984) 333
\bibitem{Bel86} {\sc J. Bellissard}
   {\em $K$-Theory of $C^*$-Algebras in Solid State Physics},
   in {\em Statistical Mechanics and Field Theory, Mathematical Aspects},
   T.C. Dorlas, M.N. Hugenholtz, M. Winnink (eds.),
   Lect. Notes in Phys. Vol. {\bf 257} (1986) 99-156, Springer Verlag,
   {\em Ordinary Quantum Hall Effect and Non-Commutative Cohomology},
   in {\em Localization in Disordered Systems}, W. Weller, P. Ziesche (eds.),
   Teubner Verlag Leipzig (1987)\\
   {\sc J. Bellissard, S. Nakamura}
   {\em Low Energy Bands do not Contribute to Quantum Hall Effect},
   Commun. Math. Phys. {\bf 131} (1990) 282-305
\bibitem{BW90} {\sc B. Block, X.G. Wen}
   {\em Effective Theories of Fractional Quantum Hall Effect at Genereic
   Filling Fractions},
   Phys. Rev. {\bf B42} (1990) 8133,
   {\em Structure of Microscopic Theory of the Quantum Hall Effect},
   Phys. Rev. {\bf B43} (1991) 8337
\bibitem{CDT93} {\sc A. Cappelli, G.V. Dunne, C.A. Trugenberger,
   G.R. Zemba}
   {\em Confomal Symmetry and Universal Properties of Quantum Hall States},
   Nucl. Phys. {\bf B398} (1993) 531-567
\bibitem{CTZ93} {\sc A. Cappelli, C.A. Trugenberger, G.R. Zemba}
   {\em Infinite symmetry in the quantum-Hall-effect},
   Nucl. Phys. {\bf B396} (1993) 465,
   {\em Classification of Quantum Hall Universality Classes by
   $\w_{1+\infty}$ Symmetry},
   Phys. Lett. {\bf 306B} (1993) 100
\bibitem{CaRa81} {\sc J.L. Cardy, E. Rabinovici}
   {\em Phase Structure of $Z_p$ Models in the Presence of a $\theta$
   Parameter},
   Nucl. Phys. {\bf B205} (1981) 1-16\\
   {\sc J.L. Cardy}
   {\em Duality and the $\theta$ Parameter in Abelian Lattice Models},
   Nucl. Phys. {\bf B205} (1981) 17-26
\bibitem{CMM91} {\sc C. Cristofano, G. Maiella, R. Musto, F. Nicodemi}
   {\em Coulomb Gas Approach to Quantum Hall Effect},
   Phys. Lett. {\bf B262} (1991) 88,
   {\em Coulomb Gas Description of the Collective States for the Fractional
   Quantum Hall Effect},
   Mod. Phys. Lett {\bf A6} (1991) 1779,
   {\em Theoretical Aspects of Quantum Hall Effect and Two-Dimensional CFT},
   Mod. Phys. Lett. {\bf A6} (1991) 2217
\bibitem{DVV88} {\sc R. Dijkgraaf, E. Verlinde, H. Verlinde}
   {\em $c=1$ Conformal Field Theories on Riemann Surfaces},
   Commun. Math. Phys. {\bf 155} (1988) 649
\bibitem{DST93} {\sc R.R. Du, H.L. Stromer, D.C. Tsui,L.N. Pfeifer,
   K.W. West}
   {\em Experimental Evidence for New Particles in the Fractional
   Quantum Hall Effect},
   Phys. Rev. Lett. {\bf 70} (1993) 2944
\bibitem{EFHHNV92} {\sc W. Eholzer, M. Flohr, A. Honecker, R. H\"ubel,
   W. Nahm, R. Varnhagen}
   {\em Representations of $\w$-Algebras with Two Generators and New
   Rational Models},
   Nucl. Phys. {\bf B383} (1992) 249
\bibitem{Flo93} {\sc M. Flohr}
   {\em $\w$-Algebras, New Rational Models and the Completeness of the
   $c=1$ Classification},
   Commun. Math. Phys. {\bf 157} (1993) 179
\bibitem{Flo94} {\sc M. Flohr}
   {\em Curiosities at Effective $c = 1$},
   Mod. Phys. Lett. {\bf A9} (1994) 1071
\bibitem{FlVa93} {\sc M. Flohr, R. Varnhagen}
   {\em Infinite symmetry in the fractional quantum-Hall-effect},
   Jour. Phys. {\bf A27} Math. Gen. (1994) 3999
\bibitem{FrKe91} {\sc J. Fr\"ohlich, T. Kerler}
   {\em Universality in Quantum Hall Systems},
   Nucl. Phys. {\bf B354} (1991) 369
\bibitem{FST94} {\sc J. Fr\"ohlich, U.M. Studer, E. Thiran}
   {\em An $ADE-O$ Classification of Minimal Incompressible Quantum Hall
   Fluids},
   preprint K.U. Leuven, May 1994, cond-mat/9406009
\bibitem{FZ91} {\sc J. Fr\"ohlich, A. Zee}
   {\em Large Scale Physics of the Quantum Hall Fluid},
   Nucl. Phys. {\bf B364} (1991) 517
\bibitem{Fub91} {\sc S. Fubini}
   {\em Vertex Operators and Quantum Hall Effect},
   Mod. Phys. Lett. {\bf A6} (1991) 347
\bibitem{Gin88} {\sc P. Ginsparg}
   {\em Curiosities at $c = 1$},
   Nucl. Phys. {\bf B295}[FS21] (1988) 153
\bibitem{Hld83} {\sc F.D.M. Haldane}
   {\em Fractional Quantization of the Hall Effect: A Hierarchy of
   Incompressible Quantum Fluid States},
   Phys. Rev. Lett. {\bf 51} (1983) 605-608,\\
   {\sc F.D.M. Haldane, E.H. Rezayi}
   {\em Periodic Laughlin-Jastrow Wave Functions for the Fractional Quantum
   Hall Effect},
   Phys. Rev {\bf B31} (1985) 2529
\bibitem{Hlp82} {\sc B.I. Halperin}
   {\em Quantized Hall Conductance, Current-Carrying Edge States and the
   Existence of Extended States in a Two-Dimensional Disordered Potential},
   Phys. Rev. {\bf B25} (1982) 2185-2190
\bibitem{Hlp84} {\sc B.I. Halperin}
   {\em Statistics of Quasiparticles and the Hierarchy of Fractional
   Quantized Hall States},
   Phys. Rev. Lett. {\bf 52} (1984) 1583-1586
\bibitem{HLR93} {\sc B.I. Halperin, P.A. Lee, N. Read}
   {\em Theory of the Half-Filled Landau Level},
   Phys. Rev. {\bf B47} (1993) 7312
\bibitem{Jai89a} {\sc J.K. Jain}
   {\em Composite Fermion Approach to the Quantum Hall Effect},
   Phys. Rev. Lett. {\bf 63} (1989) 199-202,
   {\em Incompressible Quantum Hall States},
   Phys. Rev. {\bf B40} (1989) 8079-8082
\bibitem{Jai89b} {\sc J.K. Jain}
   {\em Microscopic Theory of the Fractional Quantum Hall Effect},
   Advances in Physics {\bf 41},2 (1992) 105-146
\bibitem{Kar94} {\sc D. Karabali}
   {\em Algebraic Aspects of the Fractional Quantum Hall Effect},
   Nucl. Phys. {\bf B419} (1994) 437,
   {\em ${\cal W}_{\infty}$ Algebras in the Quantum Hall Effect},
   Nucl. Phys. {\bf B428} (1994) 531,
   {\em ${\cal W}_{\infty}$ Algebras and Incompressibility in the Quantum
   Hall Effect},
   preprint IASSNS-HEP-94/93, hepth-9411082
\bibitem{Kie91} {\sc St. Kievelson, D.-H. Lee, S.-C. Zhang}
   {\em Global Phase Diagram in the Quantum Hall Effect},
   Phys. Rev {\bf B46} (1992) 2223-2238
\bibitem{Kir89} {\sc E.B. Kiritsis}
   {\em Proof of the Completeness of the Classification of Rational
   Conformal Theories with $c=1$},
   Phys. Lett. {\bf B217} (1989) 427,
   {\em Some Proofs on the Classification of Rational Conformal Field
   Theories with $c=1$},
   California Institute of Technology Preprint CALT-68-1510 (1988)
\bibitem{Kli80} {\sc K. Klitzing, G. Dorda, M. Pepper}
   {\em New Method for High-Accuracy Determination of the Fine-Structure
   Constant Based on Quantized Hall Resistance},
   Phys. Rev. Lett. {\bf 45} (1980), 494
\bibitem{Koh85} {\sc M. Kohmoto}
   {\em Topological Invariant and the Quantization of the Hall Conductance},
   Ann. Phys. (NY) {\bf 160} (1985) 343-354
\bibitem{Lau81} {\sc R.B. Laughlin}
   {\em Quantized Hall Conductivity in Two Dimensions},
   Phys. Rev. {\bf B23} (1981), 5632-5633
\bibitem{Lau83} {\sc R.B. Laughlin}
   {\em Anomalous Quantum Hall Effect: An Incompressible Quantum Fluid with
   Fractionally Charged Excitations},
   Phys. Rev. Lett. {\bf 50} (1983), 1395-1398
\bibitem{FL90} {\sc A. Lopez, E. Fradkin}
   {\em Fractional Quantum Hall Effect and Chern-Simons Gauge Theories},
   Phys. Rev. {\bf B44} (1990) 5246
   {\em Universality in the Fractional Quantum Hall Effect},
   Nucl. Phys. {\bf B33}C Proc. Suppl. (1993) 67-91
\bibitem{Lut93} {\sc C.A. L\"utken}
   {\em Geometry of Renormalization Group Flows Constrained by Discrete
   Global Symmetries},
   Nucl. Phsy. {\bf B396} (1993) 670-692,
   {\em Global Phase Diagrams for Charge Transport in Two Dimensions},
   J. Phys. {\bf A26} Math. Gen. (1993) L811-L817,\\
   {\sc C.A. L\"utken, G.G. Ross}
   {\em Delocalization, Duality, and Scaling in the Quantum Hall System},
   Phys. Rev. {\bf B48} (1993) 2500
\bibitem{MR90} {\sc G. Moore, N. Read}
   {\em Nonabelions in the Fractional Quantum Hall Effect},
   Nucl. Phys. {\bf B360} (1991) 362
\bibitem{NTW85} {\sc Q. Niu, D.J. Thouless, Y.S. Wu}
   {\em Quantized Hall Conductance as a Topological Invariant},
   Phys. Rev. {\bf B31} (1985) 3372-3377,\\
   {\sc Q. Niu, D.J. Thouless}
   {\em Quantum Hall Effect with Realistic Boundary Conditions},
   Phys. Rev. {\bf B35} (1987) 2188-2197,\\
   {\sc D.J. Thouless, M. Kohmoto, M.P. Nightingale, M. den Nijs}
   {\em Quantized Hall Conductance in a Two-Dimensional Periodic Potential},
   Phys. Rev. Lett. {\bf 49} (1982) 405-408
\bibitem{Nov81} {\sc S.P. Novikov},
   Sov. Math. Dokl. {\bf 23} (1981) 298-303
\bibitem{PRS90} {\sc C.N. Pope, L.J. Romans, X. Shen}
   {\em The Complete Structure of $\w_{\infty}$},
   Phys. Lett. {\bf B236} (1990) 173,
   {\em $\w_{\infty}$ and the Racah-Wigner Algebra},
   Nucl. Phys, {\bf B339} (1990) 191
\bibitem{PG87} {\sc R.E.Prange, S.M.Girvin} (eds.)
   {\em The Quantum Hall Effect},
   Graduate Texts in Contemporary Physics, Springer Verlag (1987)
\bibitem{Sto91} {\sc M. Stone}
   {\em Superfluid Dynamics of the Fractional Quantum Hall State},
   Phys. Rev. {\bf B42} (1990) 212-217,
   {\em Edge Waves in the Quantum Hall Effect},
   Ann. Phys. {\bf 207} (1991) 38-52,
   {\em Schur Functions, Chiral Bosons and the Quantum Hall Effect},
   Phys. Rev {\bf B42} (1990) 8399-8404
   {\em Vertex Operators in the Quantum Hall Effect},
   Int. Jour. Mod. Phys. {\bf B5} (1991), 509,\\
   {\sc M. Stone, H.W. Wyld, R.L. Schult}
   {\em Edge Waves in the Quantum Hall Effect and Quantum Dots},
   Phys. Rev. {\bf B45} (1992) 14156
\bibitem{Sto92} {\sc M. Stone} (ed.)
   {\em Quantum Hall Effect},
   World Scientific (1992)
\bibitem{TSG82} {\sc D.C. Tsui, H.L. Stromer, A.C. Gossard}
   {\em Two-Dimensional Magnetotransport in the Extreme Quantum Limit},
   Phys. Rev. Lett. {\bf 48} (1982), 1559
\bibitem{Var94} {\sc R. Varnhagen}
   {\em Topology and Fractional Quantum Hall Effect},
   preprint BONN-TH-94-22, November 1994, hep-th/9411031
\bibitem{Ver91} {\sc E. Verlinde}
   {\em A Note on Braid Statistics and the Non-Abelian Aharanov-Bhom Effect},
   in {\em Modern Quantum Field Theory},
   World Scientific (1991) 450-461 or in Ref. \cite{Sto92} 258-269
\bibitem{Wen92} {\sc X.G. Wen}
   {\em Non-Abelian Statistics in the Fractional Quantum Hall States},
   Phys. Rev. Lett. {\bf 66} (1991) 802-805,
   {\em Chiral Luttinger Liquid and the Edge Excitations in the Fractional
   Quantum Hall States},
   Phys. Rev. {\bf B41} (1990) 12838-12844
   {\em Theory of the Edge States in Fraction Quantum Hall Effects},
   Int. Jour. Mod. Phys. {\bf B6},10 (1992) 1711-1762
\bibitem{WZ92} {\sc X.G. Wen, A. Zee}
   {\em A Classification of Abelian Quantum Hall States and Matrix
   Formulation of Topological Fluids},
   Institute of Theoretical Physics Santa Barbara Preprint NST-ITP-92-10
   (1992),\\
   {\sc E.Keski-Vakkuri, X.G. Wen}
   {\em The Ground State Structure and Modular Transformations of
   Fractional Quantum Hall States on a Torus},
   Int. J. Mod. Phys. {\bf B7} (1993) 4227-4260,\\
   {\sc X.G. Wen, E. Dagotto, E. Fradkin}
   {\em Anyons on a Torus},
   Phys. Rev. {\bf B42} (1990) 6110
\bibitem{Wil82} {\sc F. Wilczek}
   {\em Quantum Mechanics of Fractional-Spin Particles},
   Phys. Rev. Lett. {\bf 49} (1982) 957
\bibitem{WRP93} {\sc R.L. Willet, R.R. Ruel, M.A. Paalanen. K.W. West,
   L.N. Pfeiffer}
   {\em Enhanced Finite-Wave-Vector Conductivity at Multiple
   Even-Denominator Filling Factors in Two-Dimensional Electron Systems},
   Phys. Rev. {\bf B47} (1993) 7344
\bibitem{Wit83} {\sc E. Witten}
   {\em Global Aspects of Current Algebra},
   Nucl. Phys. {\bf B223} (1983) 422,
   {\em Current Algebra, Baryons, and Quark Confinement},
   Nucl. Phys. {\bf B223} (1983) 433
\bibitem{Wit89} {\sc E.\ Witten}
   {\em Quantum Field Theory and the Jones Polynomial},
   Commun. Math. Phys. {\bf 122} (1989) 351
\bibitem{Wit92} {\sc E. Witten}
   {\em Ground Ring of Two-Dimensional String Theory},
   Nucl. Phys. {\bf B373} (1992) 187-213
  } %%% end footnotesize %%%
  \end{thebibliography}
  \end{document}